\documentclass{aa}
\usepackage[english]{babel} 
\usepackage{txfonts,natbib}
\usepackage{graphicx}
\bibpunct{(}{)}{;}{a}{}{,}

\def\Mdot{\hbox{$\dot{M}$}}                     %% Mdot      
\def\Msun{\hbox{M$_{\odot}$}}               %% solar mass
\def\Rstar{\hbox{R$_{\star}$}}               %% stellar radius
\def\kms{\hbox{km$\;$s$^{-1}$}}
\def\Tstar{\hbox{T$_{\star}$}}               %% stellar temperature
\def\mic{\hbox{$\mu$m}}                     %% micron
\def\Mstar{\hbox{M$_{\star}$}}               %% stellar mass
\def\Lsun{\hbox{L$_{\odot}$}}               %% solar luminosity
\def\Lstar{\hbox{L$_{\star}$}}               %% stellar luminosity

\begin{document}

\title{Probing the mass-loss history of AGB and red supergiant stars
  from CO rotational line profiles} \subtitle{I.\ Theoretical model --
  Mass-loss history unravelled in \object{VY CMa}}

%%%%%%%%%%%%%%%%%%%%%%%%%%%%%%%%%%%%%%%%%%%%%%%%%%%%%%%%%%%%%%%%%%%%%%%%%%%

\author{L.\ Decin \inst{1,2} \thanks{\emph{Postdoctoral Fellow of the Fund for
Scientific Research, Flanders}}
\and S.\ Hony  \inst{1}
\and A.\ de Koter \inst{2}
\and K.\ Justtanont \inst{3}
\and A.G.G.M.\ Tielens \inst{4,5} \thanks{Current address: MS 245-3, NASA Ames Research Center, Moffett Field, CA 94035-1000}
\and L.B.F.M.\ Waters\inst{1,2}}

\offprints{L.\ Decin, e-mail: Leen.Decin@ster.kuleuven.ac.be}

\institute{
  Department of Physics and Astronomy, Institute for Astronomy,
  K.U.Leuven, Celestijnenlaan 200B, B-3001 Leuven, Belgium
  \and
  Sterrenkundig Instituut Anton Pannekoek, University of Amsterdam,
  Kruislaan 403, NL-1098 Amsterdam, The Netherlands
  \and
  Stockholm Observatory, AlbaNova University Center, S-106 91 Stockholm,
  Sweden
  \and
  Kapteyn Astronomical Institute, PO Box 800, NL-9700 AV Groningen, The
  Netherlands
  \and 
  SRON, PO Box 800, NL-9700 AV Groningen, The Netherlands
}

\date{17 March 2006; 9 June 2006}

%%%%%%%%%%%%%%%%%%%%%%%%%%%%%%%%%%%%%%%%%%%%%%%%%%%%%%%%%%%%%%%%%%%%%%%%%%%

\abstract
{Mass loss plays a dominant role in the evolution of low mass
  stars while they are on the Asymptotic Giant Branch (AGB). The gas
  and dust ejected during this phase are a major source in the mass
  budget of the interstellar medium. Recent studies have pointed
  towards the importance of variations in the mass-loss history of
  such objects.}
{By modelling the full line profile of low excitation CO lines emitted
  in the circumstellar envelope, we can study the mass-loss
  \emph{history} of AGB stars.}
{We have developed a non-LTE radiative transfer code, which calculates
  the velocity structure and gas kinetic temperature of the envelope
  in a self-consistent way. The resulting structure of the envelope
  provides the input for the molecular line radiative calculations
  which are evaluated in the comoving frame. The code allows for the
  implementation of modulations in the mass-loss rate. This code has
  been benchmarked against other radiative transfer codes and is shown
  to perform well and efficiently.}
{We illustrate the effects of varying mass-loss rates in case of a
  superwind phase. The model is applied to the well-studied case of
  \object{VY CMa}. We show that both the observed integrated line strengths as
  the spectral structure present in the observed line profiles,
  unambiguously demonstrate that this source underwent a phase of high
  mass loss ($\sim 3.2 \times 10^{-4}$\,\Msun\,yr$^{-1}$) some 1000\,yr ago.
  This phase took place for some 100\,yr, and was preceded by a low
  mass-loss phase ($\sim 1 \times 10^{-6}$\,\Msun\,yr$^{-1}$) taking some
  $800$\,yr. The current mass-loss rate is estimated to be in the
  order of $8 \times 10^{-5}$\,\Msun\,yr$^{-1}$.}
{In this paper, we demonstrate that both the relative strength of the
  CO rotational line profiles and the (non)-occurrence of spectral
  structure in the profile offer strong diagnostics to pinpoint the
  mass-loss history.}

\keywords{Line: profiles, Radiative transfer, Stars: AGB and post-AGB,
  (Stars): circumstellar matter, Stars: mass loss, Stars: individual:
  \object{VY CMa}}

\maketitle

%%%%%%%%%%%%%%%%%%%%%%%%%%%%%%%%%%%%%%%%%%%%%%%%%%%%%%%%%%%%%%%%%%%%%%%%%%%

\section{Introduction}
\label{introduction}
When low and intermediate mass stars approach the end of their lives
and ascend the Asymptotic Giant Branch (AGB), mass loss dominates the
subsequent evolution ultimately leading to the removal of the entire
envelope. In this process, the star develops a dust-driven wind with a
velocity of 10 -- 15\,km\,s$^{-1}$ \citep[][and references
therein]{Habing2003agbs.conf.....H}. AGB stars lose $\sim 35 - 85$ per
cent of their mass through the stellar wind before ending up as a
white dwarf \citep{Marshall2004MNRAS.355.1348M}. Likewise, more
massive stars (M$_{\rm{ZAMS}} \ga 8$\,\Msun) may pass through a red
supergiant phase and lose mass in a similar manner.

The mass-loss properties change with time as these stars ascend the
AGB and become more luminous. It is often postulated and in rare cases
observed \citep[e.g.]{Justtanont1996ApJ...456..337J,
  vanLoon2003MNRAS.341.1205V, Marshall2004MNRAS.355.1348M} that the
AGB evolution ends in a very high mass-loss phase, the so-called
superwind phase \citep{Iben1983ARA&A..21..271I}. The nature and onset
of such a superwind are far from understood.

This superwind phase is not the only type of variable mass loss.
Recent observations of (post) AGB objects and Planetary Nebulae (PNe)
show that winds from late type stars are far from being smooth. The
shell structures found around e.g. \object{CRL 2688} \citep[the Egg
Nebula,][]{ Ney1975ApJ...198L.129N, Sahai1998ApJ...493..301S},
\object{NGC 6543} \citep[the Cat's Eye
Nebula,][]{Harrington1994AAS...185.9003H} and the AGB star \object{IRC
  +10 216} \citep{Mauron1999A&A...349..203M,
  Mauron2000A&A...359..707M, Fong2003ApJ...582L..39F}, indicate that
the outflow has quasi-periodic oscillations, with density contrasts
corresponding to mass-loss variations of up to a factor
$\sim$100$-$1000 \citep{Schoier2005A&A...436..633S}. The time scale
for these oscillations is typically a few hundred years, i.e.\ too
long to be a result of stellar pulsation, which has a period of a few
hundred days, and too short to be due to nuclear thermal pulses, which
occur once in ten thousand to hundred thousand years. Presently, the
physical mechanism responsible for these variations, their rate of
occurrence and their importance in terms of the amount of ejected
matter involved are unknown. Possibly they are linked to the
hydrodynamical properties of a dust-driven wind. For example,
\citet{Simis2001A&A...371..205S} find that a feedback mechanism
between the wind-acceleration and dust growth at the base of the wind
may cause modulations in the mass-loss rate.

Already in the late seventies, low excitation rotational
carbon-monoxide (CO) emission lines emerging from the circumstellar
envelopes (CSEs) around AGBs were observed and analysed to estimate
the mass-loss rates during the AGB phase
\citep[e.g.][]{Zuckermann1977ApJ...211L..97Z,
  Morris1980ApJ...236..823M, Knapp1980ApJ...242L..25K,
  Knapp1982ApJ...252..616K, Knapp1985ApJ...292..640K}. CO has proven
to be instrumental for this since (almost) all the elemental carbon is
locked up in CO throughout most of the envelope. While previous
studies have assumed a constant mass-loss rate in the analysis of
rotational CO lines, the goal of our study is \emph{to exploit the
  rotational CO line profiles of different excitation levels as a
  complementary diagnostic to probe variations in mass-loss rate}.
With the arrival of HIFI and PACS the powerful spectrometers on the
Herschel Space Observatory --- ESA's far infrared mission to be
launched in 2008 --- and ALMA --- the submillimeter interferometer
which is being built by ESO and NRAO in Chile --- new high-quality
data of CO emission lines will put new light on this ongoing research
of understanding the geometry, physical conditions and mass-loss
history of AGB stars.

In this paper, we describe our non-LTE (non- Local Thermodynamic
Equilibrium) radiative transfer code developed to
predict the CO molecular line strengths. In
Sect.~\ref{temperature_velocity}, the equations governing the
temperature and velocity structure in the CSE are presented. Since the
CO rotational line profiles are very sensitive to the temperature
structure in the CSE, the kinetic temperature is calculated
self-consistently taking the main heating and cooling processes into
account (Sect.~\ref{temperature}). The code can treat any density
profile, allowing us to deduce the amplitude of the mass-loss
variations (Sect.~\ref{var_mass1}). The non-LTE radiative transfer
code is described in Sect.~\ref{radtrans}, and the effect of
variations in mass loss on the predicted CO profiles is studied. In
Sect.~\ref{vycma}, we apply our theoretical code to model the observed
CO rotational emission lines in the supergiant \object{VY CMa}. We
summarise the results in Sect.~\ref{summary}.

%%%%%%%%%%%%%%%%%%%%%%%%%%%%%%%%%%%%%%%%%%%%%%%%%%%%%%%%%%%%%%%%%%%%%%%%%%%

\section{Theoretical model for the temperature and velocity structure in
  the CSE}  \label{temperature_velocity}

A detailed understanding of the wind around red giants requires the
solution of the equations of hydrodynamics, specifically in the
complex region starting from the stellar surface into the region of
dust formation. The equations should include the conservation laws of
mass, momentum, and energy, in combination with a chemical model
describing the gas phase chemistry in the CSE. In the absence of such
a theory, we follow previous work, and focus on those equations
describing the velocity and temperature structure in the CSE. The main
assumption is a spherically symmetric mass loss. The code builds on
the work of \citet[][hereafter referred to as
J94]{Justtanont1994ApJ...435..852J}. Main changes are the inclusion of
new and improved heating and cooling rates (Sect.~\ref{temperature}),
and the implementation of a variable mass loss
(Sect.~\ref{var_mass1}). Helium, in addition to atomic and molecular
hydrogen, is now also included as collision partner.
  
%%%%%%%%%%%%%%%%%%%%%%%%%%%%%%%%%%%%%%%%%%%%%%%%%%%%%%%%%%%%%%%%%%%%%%%%%%%
\subsection{The gas and dust velocity}\label{velocity}
In this study, we will focus on red giants with outflow velocities in
excess of 5\,km\,s$^{-1}$, and mass-loss rates larger than $3 \times
10^{-7}$\,\Msun\,yr$^{-1}$. These winds are believed to be driven by
radiation pressure on the dust which condenses in the outer parts of
the atmosphere \citep{Winters2003A&A...409..715W}. In this situation,
the equations expressing the conservation of mass and momentum are
respectively given by \citep[e.g.][hereafter referred to as GS76 and
T83 respectively]{Goldreich1976ApJ...205..144G,
  Tielens1983ApJ...271..702T}
\begin{equation}
  \frac{dM(r)}{dt} = \Mdot(r) = 4 \pi r^2 \rho(r) v(r) \,,
\end{equation}
and 
\begin{equation}
  v(r) \frac{dv(r)}{dr} = (\Gamma(r) -1) \frac{G M_{\star}}{r^2}\,,
\label{velstructure}
\end{equation}
where \Mdot$(r)$ refers to the gas mass-loss rate at a radial distance
$r$ from the star, $\rho(r)$ is the gas density, $v(r)$ the gas velocity,
$M_{\star}$ the stellar mass, and $\Gamma(r)$ the ratio of the radiation 
pressure force on the dust to the gravitational force.

The total dust mass-loss rate, \Mdot$_d$, can be written as (J94)
\begin{eqnarray}\label{Mdotdust}
  \lefteqn{\Mdot_d (r)  =  \int{\Mdot_d(a,r)\ da} \nonumber }\\
              & = & \int{A(r)\,a^{-3.5} n_{\rm{H}}(r) \frac{4}{3} \pi a^3 \rho_s
          4 \pi r^2 [v(r) +
          v_{\rm{drift}}(a,r)]\ da} \,, 
\end{eqnarray}
where \Mdot$_d(a,r)$ represents the dust mass-loss rate of a grain of
size $a$, $n_{\rm{H}}$ the hydrogen number density, $\rho_s$ the
specific density of dust and $v_{\rm{drift}}(a,r)$ the drift velocity
of a grain of size $a$. For the dust composition, we use silicate
grains with $\rho_s = 3.3$\,g\,cm$^{-3}$
\citep{Draine1984ApJ...285...89D}. We adopt a grain-size
distribution $n_d(a,r)\,da = A(r)\,a^{-3.5} n_{\rm{H}}(r)\,da$, with a
minimum grain-size, $a_{\rm{min}}$, of 0.005\,$\mu$m and a maximum
grain-size, $a_{\rm{max}}$, of 0.25\,$\mu$m. The slope $-3.5$ is
typical for interstellar dust \citep{Mathis1977ApJ...217..425M}. It is
unclear whether this slope is representative for circumstellar dust
shells \footnote{The choice of slope has very little direct influence
on the results.  The calculated gas-to-dust ratio changes to some
extent since the absorption cross-sections are in the Rayleigh regime,
and hence scale with the total dust volume. The gas-grain drag on the
other hand scales with the geometric cross section. The grain-size
distribution also affects the temperature structure through the
gas-grain collisional heating. Since smaller grains obtain a smaller
drift velocity a steeper grain-size distribution causes less efficient
heating. On the other hand, photoelectric yields will be higher for
small grains. }.  Studies on dust condensation and growth suggest that
the circumstellar dust grain size distribution could be much steeper
\citep{Dominik1989A&A...223..227D}.  However, shattering by
grain-grain collisions could lead to a much shallower power law
\citep{Biermann1980ApJ...241L.105B}.  The quantity $A(r)$ represents
an abundance scale factor giving the number of dust particles in units
of particles per H atom. For the ISM, $A$ is a constant, estimated to
be $10^{-25.10}$\,cm$^{3.5}$/H for silicate grains and
$10^{-25.13}$\,cm$^{3.5}$/H for carbon grains
\citep{Draine1984ApJ...285...89D}. In the case of CSEs around giants,
the value of $A(r)$ may differ from the ISM value, since it
incorporates the (unknown) dust-to-gas mass ratio ($\psi$) of the CSE.
Moreover $A(r)$ varies as a function of the radius $r$ approximately
as $v(r)/(v(r)+v_{\rm{drift}}(r))$. In case of a constant mass-loss
rate, $A$ decreases rapidly at the base of the envelope w.r.t.\ the
value when the dust first condenses and the dust velocity is still
zero, as the dust is rapidly accelerated by radiation pressure,
yielding a lower dust abundance. It then increases until it reaches a
constant value when the ratio of $v(r)$ to $(v(r)+v_{\rm{drift}}(r))$
is constant. $A(r)$ is calculated from the adopted gas mass-loss rate
\Mdot\ and dust-to-gas mass ratio $\psi$ through Eq.~(\ref{Mdotdust}).

Considering the motion of grains, and equating the radiation pressure
force with the gas drag force, the drift velocity of a grain of size
$a$ can be calculated to be \citep{Kwok1975ApJ...198..583K,
  TruongBach1991A&A...249..435T}

\begin{equation}
  v_{\rm{drift}}(a,r) = v_K(a,r) \left[\sqrt{1+x(r)^2} - x(r) \right]^{1/2} \,,
  \label{Eqvdrift}
\end{equation}
with
\begin{eqnarray}
  v_K(a,r) & = & \sqrt{\frac{v(r)} {\Mdot(r) c} \int{Q_{\lambda}(a) L_{\lambda}\ d\lambda}} \nonumber \\
  x(r) & = & \frac{1}{2} \left[\frac{v_T(r)}{v_K(r)}\right]^2 \ {\rm ,\ and} \nonumber \\
  v_T(r) & = & \frac{3}{4} \left[\frac{3 k T(r)}{\mu m_{\rm{H}}}\right]^{1/2}\,,
  \nonumber
\end{eqnarray}
with $Q_{\lambda}(a)$ the extinction efficiency as derived by
\citet{Justtanont1992ApJ...389..400J} for silicates, $L_{\lambda}$ the
monochromatic stellar luminosity at wavelength $\lambda$, $v_T(r)$ the
Maxwellian velocity, $T(r)$ the temperature, $k$ the Boltzmann
constant, $\mu$ the mean molecular weight of the gas and $m_{\rm{H}}$
the mass of the hydrogen atom. We note that the J94 assumption of
$v_{\rm{drift}}(a,r) = v_K(a,r)$ holds only in the cool outer region
where thermal velocities $v_T$ are small compared to $v_{\rm{drift}}$.
The value of $v_{\rm{drift}}(a,r)$ should be lower than $\sim
20$\,\kms, at which sputtering of the dust grains starts to become
important \citep{Kwok1975ApJ...198..583K}. When the star is luminous
and goes through a phase of low mass loss, this value can be higher
than 20\,\kms\ for a certain grain-size $a$. When this occurs, dust
grains of size $a$ are destroyed,  and hence the contributions
of the dust grain of 
size $a$ to the gas-grain collisional heating, the photoelectric
heating from dust grains and the heat exchange between dust and gas
(see Sect.~\ref{temperature}) is zero.

Using the above-mentioned equations, the ratio of the drag force to
the gravitational force $\Gamma(r)$ can be written as (J94)
\begin{equation}\label{Gamma}
  \Gamma(r) = \frac {3 v(r)}{16 \pi \rho_s c G M_{\star}
    \dot{M}(r)}\int\!\!\!{\int{\frac{Q_{\lambda}(a) L_{\lambda} \dot{M}_d(a,r)}{a
        [v(r)+v_{\rm{drift}}(a,r)]}}} \,d\lambda\,da \,.
\end{equation}

%%%%%%%%%%%%%%%%%%%%%%%%%%%%%%%%%%%%%%%%%%%%%%%%%%%%%%%%%%%%%%%%%%%%%%%%%%%
\subsection{The thermal balance equation of the gas}\label{temperature}

From the first law of thermodynamics expressing the conservation of
energy of the gas, the perfect gas law and the equation of mass
conservation, the kinetic temperature of the gas in the CSE is
governed by the relation 
\begin{eqnarray}\label{tempstructure}
  \lefteqn{\frac{1}{T(r)} \frac{dT(r)}{dr}  =  -\frac{4}{3 r}
    \left(1 + \frac{\epsilon(r)}{2} \right)}  \\
  & & + \frac{2}{3} \frac{H(r)-C(r)}{k T(r) v(r) n({\rm H}_2)(r) [f_{\rm{H}}(r) + 1 +
    f_{\rm{He}}(r)(f_{\rm{H}}(r)+2)]}  \nonumber
\end{eqnarray}
where $\epsilon(r) = d\ln v(r) / d\ln r$ the logarithmic velocity
gradient, $H(r)$ the total heating rate per unit volume, $C(r)$ the
total cooling rate per unit volume, $f_{\rm{H}}(r)$ the number
fraction of atomic to molecular hydrogen i.e.\
$n$(H)(r)/$n$(H$_2)(r)$, and $f_{\rm{He}}(r)$ the helium abundance
i.e.\ $n_{{\rm He}}(r)/n_{{\rm H}}(r) = n({\rm He})(r) / (n({\rm
  H})(r) + 2n({\rm H_2})(r))$. The first term on the right hand side
of Eq.~(\ref{tempstructure}) represents the cooling due to adiabatic
expansion in case of constant mass loss. The second term represents
the balance of the heating and collision-driven radiative cooling
processes. The denominator in this expression takes the contribution
of the different collision partners H, H$_2$, and He into account.

In order to calculate the gas kinetic temperature as a function of
radius, all the major heating and cooling sources must be taken into
account. The main processes are discussed below. Newly implemented
energy sources w.r.t.\ J94 are photoelectric heating from dust grains,
heat exchange between dust and gas, and heating by cosmic rays. All
the other heating/cooling terms underwent updates mainly concerning
the used inelastic collision rates. As demonstrated by J94, cooling by
vibrational excitation of H$_2$O molecules is unimportant, since the
critical density of H$_2$O is too low in the envelope to play a role
in the energy balance. We therefore neglect this effect. We note that
we have not yet accounted for fine-structure line cooling. This
process may be important for supergiants having a strong chromosphere
\citep[e.g.\ \object{$\alpha$ Ori},][]{Rodgers1991ApJ...382..606R}. A
proper calculation of the number densities of the main atoms and
molecules requires  
the implementation of a chemical reaction network programme, which is
planned for the future.

%%%%%%%%%%%%%%%%%%%%%%%%%%%%%%%%%%%%%%%%%%%%%%%%%%%%%%%%%%%%%%%%%%%%%%%%%%%
\subsubsection{Gas-grain collisional heating}
Both gaseous and dusty material in the CSE are accelerated away from
the central star by radiation pressure. This radiation pressure mainly
acts on the dust grains, and momentum is transferred to the gas by
collisions between gas molecules and dust grains (GS76). These
collisions are the main heat source for the gas. The viscous heat
input per unit volume can be written as
\begin{equation}
  H_{{\rm gg}}(r) = n_d(a,r) \sigma_d(a) v_{\rm{drift}}(a,r) \times \frac{1}{2} \rho(r)
  v_{\rm{drift}}^2(a,r) \,,
\end{equation}
with $\sigma_d(a)$ the geometrical cross section of a dust grain of
size $a$, being $\pi a^2$. Taking the grain distribution into account,
one obtains
\begin{eqnarray}
  H_{{\rm gg}}(r) & = &\frac{\pi}{2} \times A(r) \times m_{{\rm H}} \times  n^2({\rm
    H_2})(r) \times (f_{\rm{H}}(r)+2)^2   \nonumber \\
  && \times (1+4f_{\rm{He}}(r)) \times \int a^{-1.5} v_{{\rm drift}}^3(a,r) da\,,
\end{eqnarray}
with $m_{{\rm H}}$ the mass of the hydrogen atom.  Note that viscous
heating depends strongly on the value of $v_{\rm{drift}}(a,r)$.

%%%%%%%%%%%%%%%%%%%%%%%%%%%%%%%%%%%%%%%%%%%%%%%%%%%%%%%%%%%%%%%%%%%%%%%%%%%
\subsubsection{Photoelectric heating from dust grains}\label{heatpe}
Simple physical arguments demonstrate that the impact of far
ultraviolet (UV) radiation (with wavelengths between 912 and
1200\,\AA) removes electrons from dust grains in such an efficient way
that about 4\,\% of radiative energy can be converted into kinetic
energy, heating the gas \citep{Tielens1985ApJ...291..722T}. The
calculation of the heating rate of the gas by this process is
complicated by the fact that the electron ejection rate and the
photoelectron energy depend on the charge of the dust grains, which is
a function of temperature $T$ and electron density $n_{\rm{e}}$ in the
gas. For a single size of grain, this can be incorporated into a
semi-analytical formalism \citep{deJong1977A&A....55..137D}. Dealing
with a grain size distribution complicates the computations, since
essentially all grains will attain many different ionisation stages
under typical conditions. Such single and more highly charged grains
all contribute to the photoelectric heating
\citep{Bakes1994ApJ...427..822B}.

Computations for the photoelectric ejection of electrons from a dust
grain size distribution have only been performed in the carbon-rich
case. \citet{Bakes1994ApJ...427..822B} demonstrated that the
photoelectric heating rate increases rapidly with decreasing grain
size, and ``classical'' grains ($a \approx 0.1$\,\mic) contribute
negligibly to the total heating rate. Approximately half of the total
heating rate is due to species with $a \le 15$\,\AA. The other half is
contributed by larger species up to $a \le 100$\,\AA. This decrease in
photoelectric efficiency with increasing grain size results partly
from the decrease in photoelectric yield, i.e.\ the photoemission
yield of small particles is expected to be significantly larger than
that of bulk materials, since for bulk materials the photoelectron may
lose all its excess energy in collisions with other atoms before it
reaches the surface. Furthermore, it is aided by the increase in
average grain charge with increasing grain size, which is a direct
result of the fact that the UV absorption rate increases with grain
volume, while the recombination rate depends only weakly on $a$.
Hence, the lower limit of the grain size distribution has a
considerable influence on the total photoelectric heating rate, while
the upper limit has only little effect.

The physical nature of the circumstellar dust grains around O-rich
stars is far less understood and the coupling between the far-UV
radiation field and the gas is unclear. To approximate the
photoelectric heating efficiency, we therefore used the analytical
expression as obtained by \citet{Bakes1994ApJ...427..822B}. Since this
expression was obtained for a grain-size distribution with a lower
limit of 3\,\AA, we scaled their result with a factor 0.2 to account
for our $a_{\rm{min}}$ of 50\,\AA (see their Fig.~8)
\begin{eqnarray}
  \hat{H}_{{\rm pe}}(r) & = & 0.2 \times 10^{-24} \times \frac{3 \times 10^{-2}}{1 + 2 \times 10^{-4} G_0
    \sqrt{T(r)}/n_{\rm e}(r)} \nonumber \\
  & & \times n({\rm H_2})(r) (f_{\rm{H}}(r)+2)
\label{heatpe_bakes}
\end{eqnarray} 
in erg s$^{-1}$ cm$^{-3}$, with $G_0 = 1$ the standard value for the
intensity of the incident far-UV field in units of the Habing
interstellar radiation field. This equation is obtained for a
grain-size distribution $n_d(a,r) da = A a^{-3.5} n_{\rm{H}} da$, with
$A$ being $10^{-25.16}$. The electron density $n_{\rm{e}}(r)$ is
calculated from the reaction ${\rm{CO}} \rightarrow {\rm{O + C}}
\rightarrow {\rm{O + C^+ + e^-}}$, which is known to be the main
provider of electrons in the CSE. As demonstrated by
\citet{Mamon1988ApJ...328..797M} in their Fig.~6, the abundance of
neutral C in the CSE is always less than the CO or C$^+$ abundance.

Eq.~(\ref{heatpe_bakes}) still has to be scaled with a factor $G_0
\times \exp(-\tau_{\rm{uv}})$, where $\tau_{\rm{uv}}$ represents the
optical depth of the shell against UV radiation as measured from the
outer radius, i.e.
\begin{equation}
  H_{{\rm pe}}(r) = \hat{H}_{{\rm pe}}(r) \times G_0 \times
  \exp(-\tau_{\rm{uv}}(r))\,. 
\end{equation}
The optical depth $\tau_{\rm{uv}}(r)$ is taken to be $1.8 A_v(r)$
\citep{Hollenbach1999RvMP...71..173H}. In case of an interstellar
silicate-to-gas ratio the extinction $A_v = 1.6 \times 10^{-22} \times
N_H$,
with $N_H$ the column number density of hydrogen. The constant is
scaled to the appropriate dust-to-gas mass ratio of the shell, i.e.\
$A_v = 1.6 \times 10^{-22} / 0.01 \times \psi \times N_H$.

%%%%%%%%%%%%%%%%%%%%%%%%%%%%%%%%%%%%%%%%%%%%%%%%%%%%%%%%%%%%%%%%%%%%%%%%%%
\subsubsection{Heat exchange between dust and gas}
Since gas and dust have a different temperature, heat can be exchanged
between the two species. This process can heat/cool the gas
considerably, but has almost no effect on the dust temperature. The
heating/cooling rate per unit volume in units of erg s$^{-1}$
cm$^{-3}$ may be written as \citep{Burke1983ApJ...265..223B}
\begin{eqnarray}
  \lefteqn{H_{\Delta T}(r)  =  \int{n_H(r) n_d(a,r) \sigma_d(a)\
      \sqrt{\frac{8 k T(r)}{\pi m_H}}\ 
      \overline{\alpha_T(r)}}} \\
  &  & \times 2 k [T_d(r) - T(r)] da \nonumber\\
  & = &  4 \pi k  \sqrt{\frac{8 k}{\pi m_H}} \times A(r) \times n^2({\rm
    H_2})(r)\, (f_{\rm{H}}(r)+2)^2 \times \nonumber \\
  & & \overline{\alpha_T}(r) \, \sqrt{T(r)}\, [T_d(r) - T(r)] \left(a_{min}^{-0.5} - a_{max}^{-0.5}\right)\,,
\end{eqnarray}
with $T_d(r)$ the dust temperature and $\overline{\alpha_T}(r)$ the
average accommodation coefficient \citep{Burke1983ApJ...265..223B,
  Groenewegen1994A&A...290..531G}
\begin{equation}
  \overline{\alpha_T}(r) = 0.35\,\exp\left(-\sqrt{(T_d(r)+T(r))/500}\right) + 0.1\,.
\end{equation}
Since we do not solve the radiative transfer in the dust continuum,
the dust temperature as function of the grain-size $T_d(a,r)$ is
unknown. Instead, the dust temperature is assumed to vary as
\citep[Olofsson in][]{Habing2003agbs.conf.....H}
\begin{equation}
  T_d(r) = T_{\star} (R_{\star}/2r)^{2/(4+s)}\,,
\end{equation}
with \Tstar\ and \Rstar\ the temperature and radius of the central
star. Observational data suggest that $s \approx 1$ \citep[Olofsson
in][]{Habing2003agbs.conf.....H}.

%%%%%%%%%%%%%%%%%%%%%%%%%%%%%%%%%%%%%%%%%%%%%%%%%%%%%%%%%%%%%%%%%%%%%%%%%%%
\subsubsection{Heating by cosmic rays}
The heating rate per unit volume by cosmic rays is calculated by
\citet{Goldsmith1978ApJ...222..881G}, and given by
\begin{equation}
  H_{{\rm cr}}(r) = 6.4 \times 10^{-28} \times n({\rm H_2})(r)
  (1+f_{\rm{H}}(r)/2)(1+4f_{\rm{He}}(r))\, 
\end{equation}
in units of erg s$^{-1}$ cm$^{-3}$. The uncertainty in the numerical
factor is about a factor of two.

%%%%%%%%%%%%%%%%%%%%%%%%%%%%%%%%%%%%%%%%%%%%%%%%%%%%%%%%%%%%%%%%%%%%%%%%%%%
\subsubsection{Cooling by rotational excitation of H$_2$O}\label{coolrotH2O}
Because the typical temperatures in the CSE around evolved stars are
low, radiative cooling mechanisms involving the collisional excitation
of the rotational levels of abundant molecules form the major sinks of
energy. H$_2$O (this section) and CO (next section) with their high
dissociation energy and substantial dipole moment are very important
in this respect.

It is beyond the scope of this paper to calculate the population of
the many H$_2$O rotational levels in a self-consistent manner.
Instead, we will follow GS76 in approximating the structure of a
H$_2$O molecule by a three-level system: two rotational levels in the
ground electronic vibrational state $v=0$ at energy $\sim kT$ and one
rotational level at $v=1$. Assuming that all of the pure rotational
transitions in the ground vibrational state have the same excitation
temperature $T_{\rm{exc}}$, the heat loss rate per unit volume per
unit time is given by the difference between the collisional rate for
excitation and de-excitation
 
{\small
\begin{eqnarray}
  \lefteqn{C_{{\rm H_2O, rot}}(r) = n({\rm H_2O})(r) \times h\nu_{2,1}(r)
    \times \big[n({\rm H})(r) \langle\sigma
    v\rangle_{{\rm (H-H_2O)}} +  }\nonumber \\
  & & n({\rm H_2})(r) \langle\sigma v\rangle_{{\rm (H_2-H_2O)}} +  
  n({\rm He})(r) \langle\sigma v\rangle_{{\rm (He-H_2O)}} \big] \times
  \nonumber\\
  & &  \times \big[\exp(-h\nu_{2,1}(r)/kT(r)) - \exp(-h\nu_{2,1}(r)/kT_{\rm{exc}}(r))\big]\,,
\end{eqnarray}}
\noindent with $\nu_{2,1}(r)$ the transition frequency ($\,=\,2.6
\times 10^{11} T_{\rm{exc}}^{0.5}(r)$ s$^{-1}$), and $\langle \sigma v
\rangle(r)$ the inelastic collision rate constant. Quite often the
He-H$_2$O inelastic collisional rate constant is assumed to be a
factor $\sqrt{2}$ lower than the H$_2$-H$_2$O rate constant due to the
difference in mass \citep[e.g.][]{Groenewegen1994A&A...290..531G}.
However, as demonstrated by \citet{Philips1996ApJS..107..467P},
excitation by para-H$_2$ is not too different from excitation by He
atoms at excitation temperatures from 20 to 140\,K, with most rates
being within a factor 1 -- 3 larger, but excitation by ortho-H$_2$ is
significantly different, with some rates an order of magnitude larger
than rates for excitation by He atoms. From Tables 6 and 7 in
\citet{Philips1996ApJS..107..467P}, we use as mean values for the
ratio of the inelastic collision rates {\small{
\begin{eqnarray}
  \left|\frac{\langle \sigma v \rangle_{{\rm (H_2(j=0) - H_2O_{para})}}}{\langle
      \sigma v \rangle_{{\rm (He - H_2O_{para})}}}\right| \approx 1.7 &  &
  \left|\frac{\langle \sigma v \rangle_{{\rm H_2(j=0) - H_2O_{ortho}}}}{\langle
      \sigma v \rangle_{{\rm (He - H_2O_{ortho})}}}\right| \approx 1.6 \nonumber \\
  \left|\frac{\langle \sigma v \rangle_{{\rm H_2(j=1) - H_2O_{para}}}}{\langle
      \sigma v \rangle_{{\rm (He - H_2O_{para})}}}\right| \approx 7.7 &  &
  \left|\frac{\langle \sigma v \rangle_{{\rm H_2(j=1) - H_2O_{ortho}}}}{\langle
      \sigma v \rangle_{{\rm (He - H_2O_{ortho})}}}\right| \approx 7.8\,. \nonumber
\end{eqnarray}}}
Theoretical rate constants among the lowest 45 para and 45
ortho rotational levels of water in collisions with He atoms have been
calculated by \citet{Green1993ApJS...85..181G}. From their Tables 2 and
3, we derive
\begin{eqnarray}
  \langle \sigma v \rangle_{{\rm (He-H_2O_{para})}} & \approx & 0.21 \times
  \sqrt{T(r)} \times 10^{-11}\ {\rm cm^3\,sec^{-1}} \nonumber \\
  \langle \sigma v \rangle_{{\rm (He-H_2O_{ortho})}} & \approx & 0.13 \times
  \sqrt{T(r)} \times 10^{-11}\ {\rm cm^3\,sec^{-1}}\,. \nonumber 
\end{eqnarray}
The H-H$_2$O inelastic rate constant is assumed to be a factor 1.16
larger than the He-H$_2$O constant (this value takes both the smaller
mass and the smaller cross section of H into account). With an
ortho-to-para ratio of 3:1 for both H$_2$ and H$_2$O, we obtain as
cooling rate for rotational excitation of H$_2$O

{\small
\begin{eqnarray}
  \lefteqn{C_{{\rm H_2O, rot}}(r) = n({\rm H_2})(r)\  n({\rm
      H_2O})(r) \   h\nu_{2,1}(r) \big[0.21
    \times 10^{-11} \times }\nonumber\\
  & &  \sqrt{T(r)} \big] \times \big[0.83f_{\rm{H}}(r) + 0.715
  f_{\rm{He}}(r)\{f_{\rm{H}}(r)+2\}   + 4.5] \times \nonumber\\
  & &\times \big[\exp\{-h\nu_{2,1}(r)/kT(r)\} - \exp\{-h\nu_{2,1}(r)/k T_{\rm{exc}}(r)\}\big]\,.
\end{eqnarray}}
The excitation temperature $T_{\rm{exc}}(r)$ can be calculated from the rate
equations, using an escape probability formalism to describe the
radiative transfer in the line; see Eq.~(11) in J94.\\

The water molecules in the CSE are photodissociated by interstellar
ultraviolet photons with $\lambda \approx 1650$\,\AA. In particular,
the photochemical chain is given by ${\rm H_2O \rightarrow OH + H
  \rightarrow O + H + H}$ \citep[GS76;][]{Huggins1982AJ.....87.1828H,
  Netzer1987ApJ...323..734N}. This photodissociation of H$_2$O is a
significant source of OH molecules in the outer envelope. Since no
functional form exists for the distribution of water in the CSE ---
and the implementation of a chemical reaction network programme is
planned for the future --- we assume a similar description for the
H$_2$O abundance profile as derived by
\citet{Mamon1988ApJ...328..797M} for CO (see next section,
Eq.~(\ref{mamon})). In this functional form, we have to provide the
position $r_{1/2}^{{\rm(H_2O)}}$ where the abundance of water relative
to H$_2$ has decreased by a factor of two relative to the value at the
stellar surface. This value is set by
\citep{Groenewegen1994A&A...290..531G}
\begin{eqnarray}
  r_{1/2}^{{\rm(H_2O)}} & = &35\,\left({\overline{\dot{M}}} / 10^{-5}
    \,{\rm{M_{\odot}\,yr^{-1}}} \right)^{0.7}\nonumber\\
  & & \times (v(r)/15\,
  {\rm{km\,s^{-1}}})^{-0.4} 10^{15} \ {\rm{cm}}\,,
\end{eqnarray}
where (in case of variable mass loss) ${\overline{\Mdot}}$ is taken as
the mean of \Mdot$(r)$. This expression is derived from the results of
\citet{Netzer1987ApJ...323..734N}, who determined the OH peak number
densities in the CSEs around evolved stars. The thus assumed H$_2$O
distribution is similar to the results of \citet[][see their
Fig.~1]{Nejad1988MNRAS.230...79N}.

%%%%%%%%%%%%%%%%%%%%%%%%%%%%%%%%%%%%%%%%%%%%%%%%%%%%%%%%%%%%%%%%%%%%%%%%%%%
\subsubsection{Cooling by rotational excitation of CO}\label{coolrotCO}
\citet{Justtanont1994ApJ...435..852J} included CO cooling in a similar
way as H$_2$O rotational cooling. The H-CO inelastic rate constant is
assumed to be a factor 1.16 larger than the He-CO inelastic rate
constant due to the difference in mass (velocity) and cross section.
Using \citep{Schinke1985ApJ...299..939S}
\begin{equation}
  \langle \sigma v \rangle _{{\rm (CO-H_2)}} \approx 4 \times 10^{-12} \times
  \sqrt{T(r)}\  {\rm cm^3 s^{-1}}\,,
\end{equation}
and
\begin{equation}
  \langle \sigma v \rangle _{{\rm (CO-He)}} \approx 4 .5\times 10^{-13} \times
  \sqrt{T(r)} \  {\rm cm^3 s^{-1}}\,,
\end{equation}
\citep{McKee1982ApJ...259..647M}, we obtain
\begin{eqnarray}
  \lefteqn{C_{{\rm CO, rot}}(r) = n({\rm H_2})(r)\times n({\rm CO})(r)
    \times  h\nu_{2,1}(r) } \\ 
  & & \times \left[4.5 
    \times 10^{-13} \times \sqrt{T(r)} \right]  \nonumber \\
  & &   \times \left[1.16 f_{\rm{H}}(r) +  f_{\rm{He}}(r)\{f_{\rm{H}}(r)+2\}
    + 10 \right]  \nonumber\\
  & & \times \left[\exp\{-h\nu_{2,1}(r)/kT(r)\} -
    \exp\{-h\nu_{2,1}(r)/kT_x(r)\} \right]\,.\nonumber 
\end{eqnarray}
\noindent with $\nu_{2,1}(r) = 4.89 \times 10^{10}
T_x^{0.5}(r)$\,s$^{-1}$, and $T_x(r)$ the excitation temperature for
CO molecules defined in a similar manner as H$_2$O (see previous
section).

An important quantity is the CO abundance in the photosphere, which is
determined by the carbon and oxygen supply. During the first dredge-up
on the red giant branch the carbon abundance is depleted to roughly
two-thirds of the main sequence value while the oxygen abundance
remains almost unchanged. On the AGB, carbon may be added to the
envelope due to the third dredge-up. Typical CO and H$_2$O abundances
at the base of the CSE are $f_{\rm{CO}}(R_{\star}) =
n({\rm{CO}})(R_{\star})/n({\rm{H_2}}(R_{\star})) = 5.3 \times 10^{-4}$
and $f_{\rm{H_2O}}(R_{\star}) =
n({\rm{H_2O}})(R_{\star})/n({\rm{H_2}}(R_{\star})) = 1.2 \times
10^{-3}$ when the star arrives on the AGB. If the C/O ratio in an AGB
star has increased to 0.7, the abundances changes to $1.2 \times
10^{-3}$ and $5.1 \times 10^{-4}$ for CO and H$_2$O respectively
\citep{Groenewegen1994A&A...290..531G}.
\citet{Knapp1985ApJ...292..640K} arrived at a value of
$f_{\rm{CO}}(R_{\star}) = 3 \times 10^{-4}$ for O-rich giants, $6
\times 10^{-4}$ for S-type stars, and $8 \times 10^{-4}$ for C-rich
giants; \citet{Zuckerman1986ApJ...304..394Z} used $5 \times 10^{-4}$
for O-rich and $1 \times 10^{-3}$ for S- and C-rich giants. This
uncertainty in $f_{\rm{CO}}$ translates in an uncertainty on the
derived mass-loss rate. In our code the carbon and oxygen abundances
at the stellar radius $R_{\star}$ are free input parameters, and can
be adjusted according to the evolutionary state of the target under
study. If no information can be found on the carbon and oxygen
abundances in the outer atmosphere of the target, we opt to use the
cosmic abundance values of \citet{Anders1989GeCoA..53..197A}.

CO will be dissociated by the interstellar UV field, and its abundance
will decrease outward. \citet{Mamon1988ApJ...328..797M} have
investigated this effect on the CO abundance distribution. They
obtained that the CO abundance $f_{\rm{(CO)}}(r)$, relative to the
value at the photosphere $f_{\rm{(CO)}}(\Rstar)$, can be approximated
by
\begin{equation}
  f_{\rm{(CO)}}(r) = f_{\rm{(CO)}}(\Rstar) \exp\left[ -\ln
    2\left(r/r_{1/2}^{{\rm(CO)}}\right)^{\alpha} \right]\,,
\label{mamon}
\end{equation}
where $r_{1/2}^{{\rm(CO)}}$ denotes the position where the number
density has decreased by a factor of two. The values for
$r_{1/2}^{{\rm(CO)}}$ and $\alpha$ depend on the gas velocity and mass
loss rate and are tabulated by \citet{Mamon1988ApJ...328..797M}.

% %%%%%%%%%%%%%%%%%%%%%%%%%%%%%%%%%%%%%%%%%%%%%%%%%%%%%%%%%%%%%%%%%%%%%%%%%%%

\subsubsection{Cooling by vibrational excitation of H$_2$}

Cooling by vibrational excitation of H$_2$ molecules is simpler to
calculate than for H$_2$O. The reasons are: (1) The vibrational energy
level spacing is large ($\sim 0.6$\,eV). Hence only collisional
excitation of the first excited vibrational state needs to be
considered. (2) Since H$_2$ is dipole inactive, resulting in small
Einstein $A$-values, the lines are optically thin even in the dense
part of the CSE. (3) H$_2$ is in LTE through a large part of the
envelope (GS76, J94). The description of the heat loss per unit volume
is taken from GS76, and can be summarised as
\begin{equation}
  C_{{\rm H_2}} = A_{1,0}\,h\nu_{1.0}\,n_1(r)\,,
\end{equation}
where $A_{1,0} \approx 3\times 10^{-7}$\,s$^{-1}$ is the spontaneous
emission rate from the first vibrational state, $h\nu_{1,0} \approx
0.6$\,eV is the energy of the emitted photon, and $n_1 =
n_{\rm{H_2}}(v=1)$ the number density of vibrationally excited H$_2$
molecules. The number density $n_1$ of vibrationally excited H$_2$ by
collisions with H$_2$ and H atoms, can be written as
\begin{equation}
  n_1(r) = n({\rm H_2})(r) \frac{\alpha_n(r) \times
    \exp(-h\nu_{1,0}/kT(r))} {\alpha_n(r) \times \left[ 1 +
      \exp(-h\nu_{1,0}/kT(r)) \right] +A_{1,0}}
\end{equation}
with
\begin{equation}
  \alpha_n(r) =  n({\rm H})(r) \langle \sigma^*  v_{th}\rangle_{{\rm (H-H_2)}}
  +  n({\rm H_2})(r) \langle \sigma^* v_{th}\rangle_{{\rm
      (H_2-H_2)}}
\end{equation}
where $\langle \sigma^* v_{th}\rangle_{{\rm (H-H_2)}}$ and $\langle
\sigma^* v_{th}\rangle_{{\rm (H_2-H_2)}}$ are the collisional
de-excitation rate constants.

\noindent Theoretical values for $\langle \sigma^* v_{th}\rangle_{{\rm
    (H-H_2)}}$ are approximated by
\citep{Hollenbach1979ApJS...41..555H}
\begin{equation}
  \langle \sigma^* v_{th}\rangle_{{\rm (H-H_2)}} = 1.0 \times 10^{-12} T(r)^{1/2}
  \times \exp\left[-(1000/T(r))\right]
\end{equation}
cm$^3$ s$^{-1}$, and for $\langle \sigma^* v_{th}\rangle_{{\rm
    (H_2-H_2)}}$ \citep{Hollenbach1989ApJ...342..306H}
\begin{eqnarray}
  \langle \sigma^* v_{th}\rangle_{{\rm  (H_2-H_2)}} & = & 1.4 \times
  10^{-12} T(r)^{1/2} \nonumber \\
  & & \times \exp\left\{-[18100/(T(r) + 1200)]\right\}
\end{eqnarray}
cm$^3$ s$^{-1}$.

%%%%%%%%%%%%%%%%%%%%%%%%%%%%%%%%%%%%%%%%%%%%%%%%%%%%%%%%%%%%%%%%%%%%%%%%%%%
\subsection{Treatment of the boundary conditions}
The differential equations governing the temperature and velocity
structure (Eqs.~(\ref{velstructure}) and (\ref{tempstructure})) are
solved simultaneously. For the central star, we use a blackbody with a
user-defined temperature \Tstar. As inner boundary for the velocity
structure, we assume that the flow velocity of the gas is equal to the
local sound velocity, $v_s$, at the radius of the CSE,
$R_{\rm{inner}}$, where the dust condenses. Since many questions still
exist on the complex structure between $R_{\star}$ and
$R_{\rm{inner}}$ we assume that the temperature in this regime is
described by a power law
\begin{equation}
  T(r) = T_{\star} \left(\frac{R_{\star}}{r}\right)^{\zeta}
\end{equation}
with $\zeta \approx 0.5$, which holds when both emission and
absorption are optically thin. The velocity law can be approximated by
the classical $\beta$-law
\begin{equation}
  v(r) \simeq v_{\infty} \left(1 - \frac{R_0}{r}\right)^{\beta}
\end{equation}
with
\begin{equation}
  R_0 = R_{\star} \left\{1 - \left(\frac{v_0}{v_{\infty}}\right)^{1/\beta} \right\}
\end{equation}
with $v_0$ the velocity at the photosphere, which can be calculated
directly from $v_s$, and $\beta$ being $1/2$, a typical value in case
of winds of cool stars \citep{Schutte1989ApJ...343..369S}. Since we
are studying low excitation rotational CO line profiles being formed
quite far in the envelope, the exact value of $\epsilon$ and $\beta$
does not influence the calculated line profiles. The outer radius is
determined by the CO abundance, and is set at the radius where the CO
abundance $f_{{\rm CO}}(r)$ drops to 1\,\% of its value at the
photosphere. Adopting a value for the gas mass-loss rate, the
dust-to-gas ratio is varied until the observed terminal velocity is
obtained. Since $\Gamma(r) \propto \psi \cdot Q_{\lambda}(a)$
(Eq.~\ref{Gamma}), the derived dust-to-gas ratio $\psi$ should be
interpreted in relation to the used tabulated extinction efficiencies
$Q_{\lambda}(a)$ derived by \citet{Justtanont1992ApJ...389..400J} for
silicate grains.

The sudden onset of some heating or cooling terms, yields a change in
going from a non-stiff to a stiff set of differential equations (or
vice versa), which a standard Runge-Kutta method is not able to
handle. We therefore chose to use the DLSODA package as provided by
ODEPACK\footnote{http://www.netlib.org/odepack/ }. Going from the
inner boundary to the outer boundary, this solver automatically
selects between non-stiff and stiff methods. It uses Adams methods
(predictor-corrector) in the non-stiff case, and Backward
Differentiation Formula methods (the Gear methods) in the stiff case.
We require an absolute accuracy of 1\,K in temperature and 100\,cm\,s$^{-1}$
in velocity, i.e.\ the estimated local error in the parameter is less
than the above-mentioned accuracy.

%%%%%%%%%%%%%%%%%%%%%%%%%%%%%%%%%%%%%%%%%%%%%%%%%%%%%%%%%%%%%%%%%%%%%%%%%%%
\subsection{Variable mass loss}\label{var_mass1}
It is beyond the scope of this project to implement a full
hydrodynamical code to predict variabilities in the wind structure. We
simulate the effect of various types of variable mass loss, as e.g.\
the superwind phase or ``quasi-periodic variations'' discussed in
Sect.~\ref{introduction}, in an empirical manner by adapting the
mass-loss rate as a function of radius, i.e. $\dot{M}(r)$.

In order to incorporate a variable mass loss, we calculate for each
radial point $r$ the expected temperature and velocity as if resulting
from a constant mass outflow with the corresponding mass loss rate. In
practise this means that we solve Eqs.~(\ref{velstructure}) and
(\ref{tempstructure}) from the inner edge $R_{\rm{inner}}$ up to $r$
with a fixed \Mdot\ chosen to correspond to \Mdot($r$). In this way we
assume that the main influence on the temperature is set by the
expansion of the flow and the local heating and cooling terms. In this
approach the influence of the neighbouring shells, which may
correspond to different mass-loss rates, is neglected.

If the mass-loss behaviour is changed, the drift velocities will be
affected. This in turn causes a change in the flow velocity $v(r)$ of
the gas. This feedback is self-consistently accounted for in the
calculations of the heating and cooling rates. However, we do not take
these modified velocities into account when solving the radiative
transfer equation,  because our 
theoretical code solves the radiative transfer using the comoving
frame (CMF) method as developed by \citet[][see next
section]{Mihalas1975ApJ...202..465M} which demands that the velocity
should increase continuously with radius. In case of variable mass
loss, we use the velocity structure as inferred from the case with
constant \Mdot. For instance, in the test case of \object{GX Mon} described in
Sect.~\ref{discussion1}, the velocity structure as in the absence of a
superwind is used.

Several options are built in the code to define \Mdot$(r)$, being
constant, a heavy-side function, a block function, a picket-fence
function or analogous functions with smoother derivatives.

%%%%%%%%%%%%%%%%%%%%%%%%%%%%%%%%%%%%%%%%%%%%%%%%%%%%%%%%%%%%%%%%%%%%%%%%%%%
\subsection{Discussion}
\label{discussion1}
\begin{table}[!thp]
  \caption{\label{param_standard} Parameters of \object{GX Mon} as derived by J94.}
  \begin{tabular}{ll}\hline \hline
    \rule[-3mm]{0mm}{8mm} parameter & value \\
    \hline
    \Tstar & 2500\,K \\
    \Rstar & $4 \times 10^{13}$\,cm \\
    \Mstar & 1\,\Msun \\
    R$_{\rm{inner}}$  & 6.25\,\Rstar \\
    \Mdot & $7.2 \times 10^{-6}$\,M$_{\odot}$\,yr$^{-1}$ \\
    v$_{\rm{stoch}}$ & 1\,km\,s$^{-1}$ \\
    distance & 740\,pc \\
    \rule[-3mm]{0mm}{3mm}v$_{\infty}$ & 19 km\,s$^{-1}$ \\
    \hline
  \end{tabular}
\end{table}
\begin{figure*}[!thp]
  \includegraphics[height=\textwidth,angle=90]{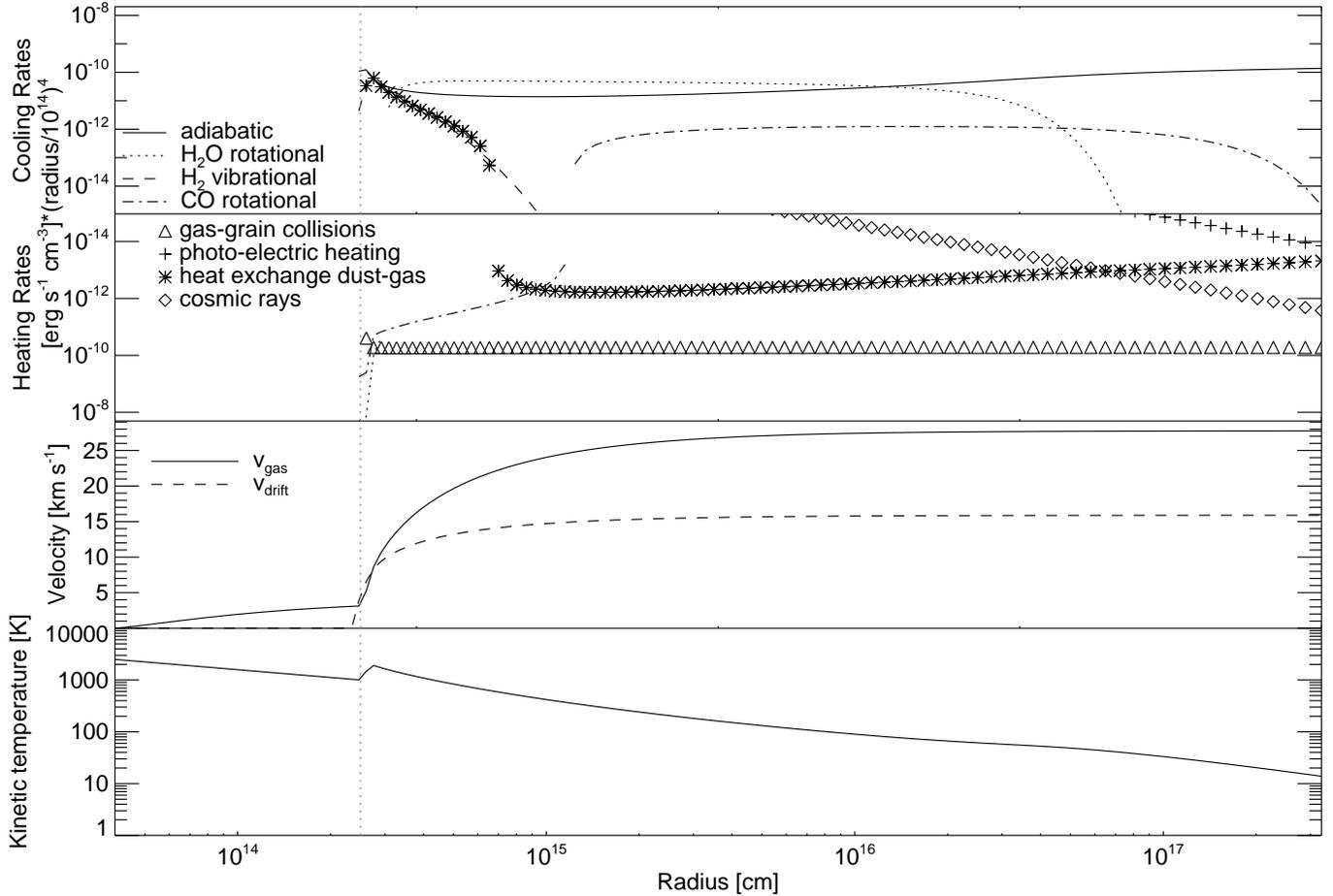}
  \caption{\label{graph_constant}The velocity structure and gas
    kinetic temperature from calculating the energy sources and sinks
    in the CSE of \object{GX Mon}. The start of the dusty envelope,
    R$_{\rm{inner}}$, is indicated by the dotted line. The drift
    velocity is displayed for the maximum grain-size, $a =
    0.25$\,$\mu$m. Note that for grains with radius smaller than the
    wavelength, $Q \propto a$; hence $v_{\rm{drift}} \propto a^{0.5}$.
    \emph{First panel:} Cooling rates; \emph{second panel:} heating
    rates; \emph{third panel:} velocity structure; \emph{fourth
      panel:} gas kinetic temperature.}
\end{figure*}
\begin{figure*}[!thp]
  \includegraphics[height=\textwidth,angle=90]{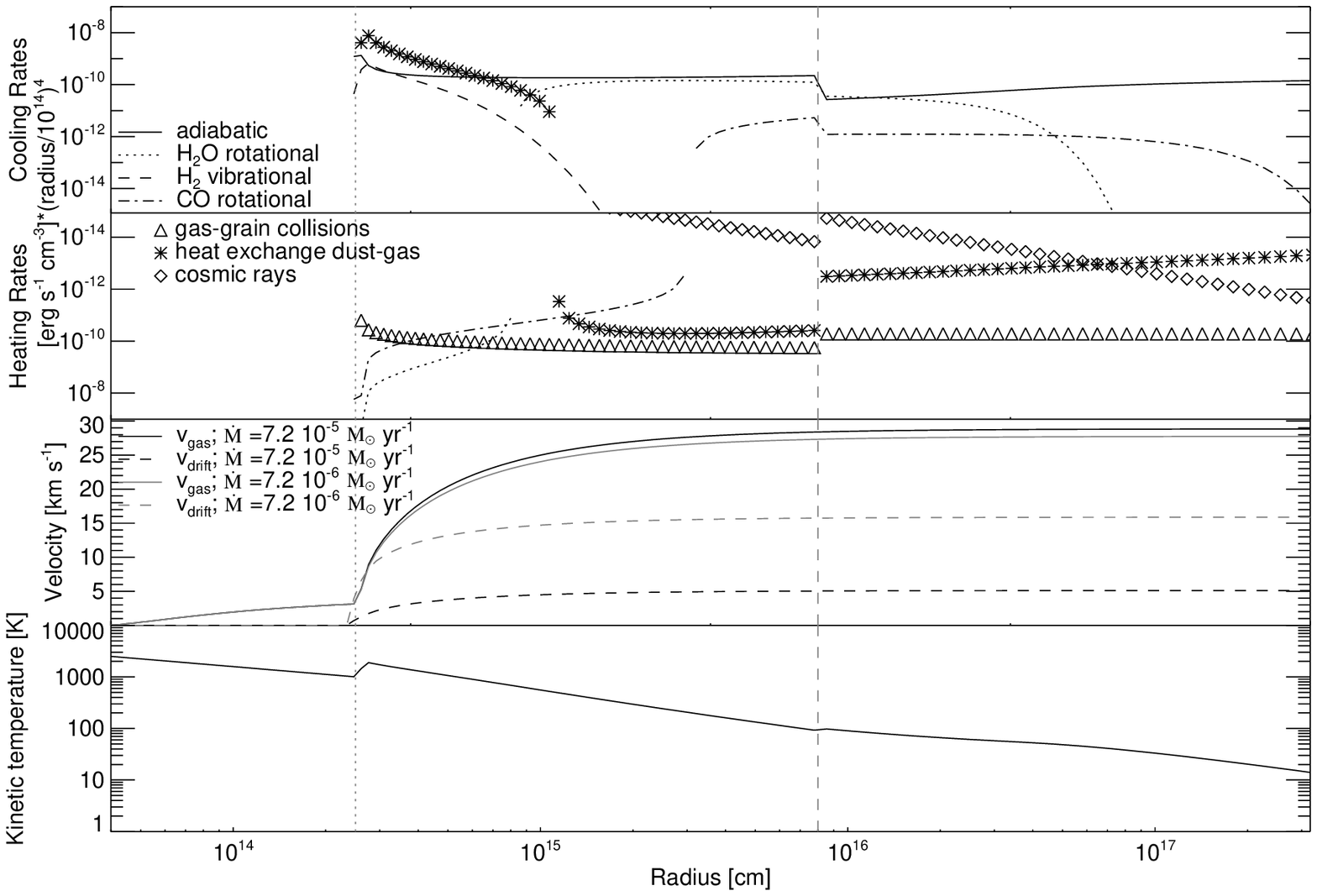}
  \caption{Same as Fig.~\ref{graph_constant}, but now with a superwind
    phase, i.e.\ with a mass-loss rate being a factor of 10 higher up
    to 200\,\Rstar\ (indicated by the dashed line). The velocity
    structure displayed in Fig.~\ref{graph_constant} is also shown in
    grey in the third panel. More details are given in the text.}
  \label{graph_2step_high}
\end{figure*}

To check the new input of heating and cooling rates, and new numerical
routines we simulated the results as obtained by J94 for the Mira-type
star \object{GX Mon}. Parameters as derived by J94 are listed in Table
\ref{param_standard}. Our final temperature and velocity structure are
displayed in Fig.~\ref{graph_constant} in the lower right and upper
right panel respectively, and should be compared with Fig.~1 in J94.

Inspecting the gas and drift velocity structure shows quite a
difference between both results. The reason for this difference is a
mistake in the implementation of the results of the ordinary
differential equation (ODE) operator (being a Runge-Kutta method with
adaptive stepsize control) in the version of J94. J94 erroneously
assigned the computed parameters being the velocity $v(r1)$,
temperature $T(r1)$ and the respective derivatives $dv(r1)/dr$ and
$dT(r1)/dr$ at radius point $r1$ also to the next radius point $r2$ in
the grid, which input was then used to calculate the parameters at the
newly derived radius point $r2'$, computed using adaptive stepsize
control. This inaccuracy yielded in case of \object{GX Mon} a terminal velocity
being $\sim 8$\,\kms\ too low and a terminal drift velocity being $\sim
5$\,\kms\ too low. Moreover, our improved treatment of the drift
velocity (Eq.~\ref{Eqvdrift}, Sect.~\ref{velocity}) lowers for \object{GX Mon}
the terminal drift velocity by 0.5\,\kms\ and the terminal flow
velocity by 2\,\kms\ compared to the approximation of
$v_{\rm{drift}}(a,r) = v_K(a,r)$ (J94).

That there is some difference between the derived temperature
structures does not come as a surprise, since new heating terms are
included and some of the ones already accounted for underwent major
updates.
\begin{itemize}
\item{Heat exchange between dust and gas and cosmic ray heating have
    been included, but they only have a minor contribution to the
    total heating rate. }
\item{The gas-grain collisional heating rate as given by J94 is
    somewhat too low due to a numerical mistake in the implementation
    of the fractional abundance of atomic hydrogen $f_{\rm{H}}$
    yielding too low a scaling factor $A$. Since this frictional
    heating is the main heating source over most of the envelope
    regime, our calculated $T(r)$ is in general higher than in J94.}
\item{The main differences in the cooling rates come from the updates
    of the inelastic collisional rates involved in the cooling by
    rotational excitation of H$_2$O and CO, being lower than in J94.}
\end{itemize} 

As an illustration of the effects of variations in the mass-loss rate,
we simulate a superwind phase with an enhanced mass loss of a factor
of ten till 200\,\Rstar\ (\,=\,$8 \times 10^{15}$\,cm) compared to the
parameters of the base-line model of \object{GX Mon} given in
Table~\ref{param_standard}. We do not attempt to better explain the
observed line profiles of \object{GX Mon}. In Sect.~\ref{vycma} we will present
the results of a parameter study focusing on explaining the
observations of \object{VY CMa}.

Fig.~\ref{graph_2step_high} displays the temperature, velocity,
heating and cooling terms for this simulation. As explained in
Sect.~\ref{var_mass1}, the velocity structure used in the radiative
transfer calculations is not coupled to a change in mass-loss rate,
but is kept as inferred from a case with constant \Mdot. In
the panel giving the velocity structure, the black lines represent the
gas and drift velocity in case \Mdot$= 7.2 \times 10^{-5}$\,\Msun\,yr$^{-1}$
would have been used to calculate the velocity structure used in the
radiative transfer calculations, the blue (grey) line is for \Mdot$=
7.2 \times 10^{-6}$\,\Msun\,yr$^{-1}$. An increase in \Mdot\ yields (1) a
decrease of the drift velocity ($v_{\rm{drift}} \propto
{\Mdot}^{-1/2}$, Eq.~(\ref{Eqvdrift})) since the coupling between dust
and gas increases, and (2) a higher terminal velocity, since there is
more dust to intercept the radiation and hence to accelerate the gas
by transfer of momentum. Which constant \Mdot\ value to choose in this
case can be debated, but as illustrated in
Fig.~\ref{graph_2step_high}, the difference between both
$v(r)$-structures is small. A more correct gas velocity profile should
display a decrease of the flow velocity from 200\,\Rstar\ (\,=\,$8
\times 10^{15}$\,cm) on.

The change in viscous heating is not dominated by the decrease in
drift velocity, but by the increase in dust and H$_2$ particles,
yielding a higher gas-grain collisional heating rate out to
200\,\Rstar. Other energy rates are increased too, with the total
influence being quite complex. In general, the temperature is higher
till 200\,\Rstar\ w.r.t.\ the constant mass-loss case displayed in
Fig.~\ref{graph_constant}. The effects of this superwind phase on the
line profiles are discussed in Sect.~\ref{var_mass2}.
 
%%%%%%%%%%%%%%%%%%%%%%%%%%%%%%%%%%%%%%%%%%%%%%%%%%%%%%%%%%%%%%%%%%%%%%%%%%%
\section{Line radiative transfer}
\label{radtrans}
Late type giants and supergiants have a ratio of stochastic to
terminal velocity that may be in the order of magnitude of $\sim 0.5$
or even 1 \citep{Che1983A&A...126..225C} yielding rather broad
scattering zones and thus a highly non-local radiative transfer.
Therefore the condition for using the Sobolev approximation
\citep{Sobolev1947, Castor1970MNRAS.149..111C} is not fulfilled
anymore. \citet{Schonberg1985A&A...148..405S} demonstrated that
non-local scattering effects cause severe errors in the source
function when using the Sobolev approximation (see also J94).

Moreover, as demonstrated by \citet{Schonberg1988A&A...195..198S},
part of the line profile of the low-excitation CO transitions is
formed under non-LTE circumstances. The decoupling radius (i.e.\ that
distance from the star where the excitation temperature in the lines
is by ten percent smaller than the kinetic temperature) is a function
of the specific line. The higher the excitation level, the smaller the
decoupling radius. Clearly, the decoupling radius decreases with
decreasing mass-loss rate. For a model with parameters almost
identical to the ones listed in Table~\ref{param_standard},
\citet{Schonberg1988A&A...195..198S} demonstrated that the excitation
of the rotational $^{12}$CO lines in the vibrational ground state is
dominated by collisions in the mass-loss range \Mdot\,$\ga 3 \times
10^{-5}$\,\Msun\,yr$^{-1}$.

We therefore opt to calculate the CO rotational line profiles using a
non-LTE radiative transfer code based on the multilevel approximate
Newton-Raphson (ANR) operator as derived by
\citet{Schonberg1986A&A...163..151S}. The ANR operator has been
derived from the comoving frame equation of radiation transfer
\citep{Mihalas1975ApJ...202..465M}. The method is similar to the
partial linearision approach. The ANR operator is a diagonal
(therefore local) operator, and can be derived with only a slight
increase in computational effort over that required to find the formal
solution. When solving for multilevel line formation in expanding
envelopes, \citet{Schonberg1986A&A...163..151S} have shown that
convergence was reached within 10 -- 20 iterations for winds around
cool giants. We note that although the method is proven to be stable,
\citet{Hillier1990A&A...231..116H} demonstrated that at very high
optical depths the diagonal ANR operator may give rise to damped,
depth-coupled oscillations, which may be due to the inability of the
diagonal operator to correctly handle the diffusion approximation. He
has shown that in these cases tridiagonal (or pentadiagonal)
Newton-Raphson operators have a superior convergence rate.

A description of the radiative transfer equations can be found in
\citet{Schonberg1988A&A...195..198S}. The calculations include
excitation by collisions with H$_2$ as well as scattering of infrared
and microwave radiation emergent from warm dust particles, from the
stellar photosphere and the cosmic background radiation. The
implementation of the dust opacity was modified to include the grain
opacity of silicates as computed in
\citet{Justtanont1992ApJ...389..400J}.

After solving the radiative transfer, line profiles for each
transition are calculated by ray-tracing using formal integration
\citep{Schonberg1988A&A...195..198S}. Writing the calculated flux as
$F_{\nu}$, the main beam temperature $T_{\rm{MB}}$ is calculated by
\citep{Schonberg1988A&A...195..198S}
\begin{equation}
  T_{\rm{MB}} = F_{\nu} \frac{\pi D^2_{\rm{TEL}}}{8 k}\,,
\end{equation}
with $D_{\rm{TEL}}$ the diameter of the antenna. Note that the line
profile will not be symmetric since the opacity in the blue wing is
produced at slightly larger $r$ than in the red wing. This kind of
asymmetry in double-peaked (optically thin) line profiles will be
stronger for resolved envelopes, and provides a useful diagnostic for
obtaining the stochastic velocity
\citep[see][]{Schonberg1988A&A...195..198S}.
\begin{figure*}[!thp]
  \includegraphics[width=\textwidth]{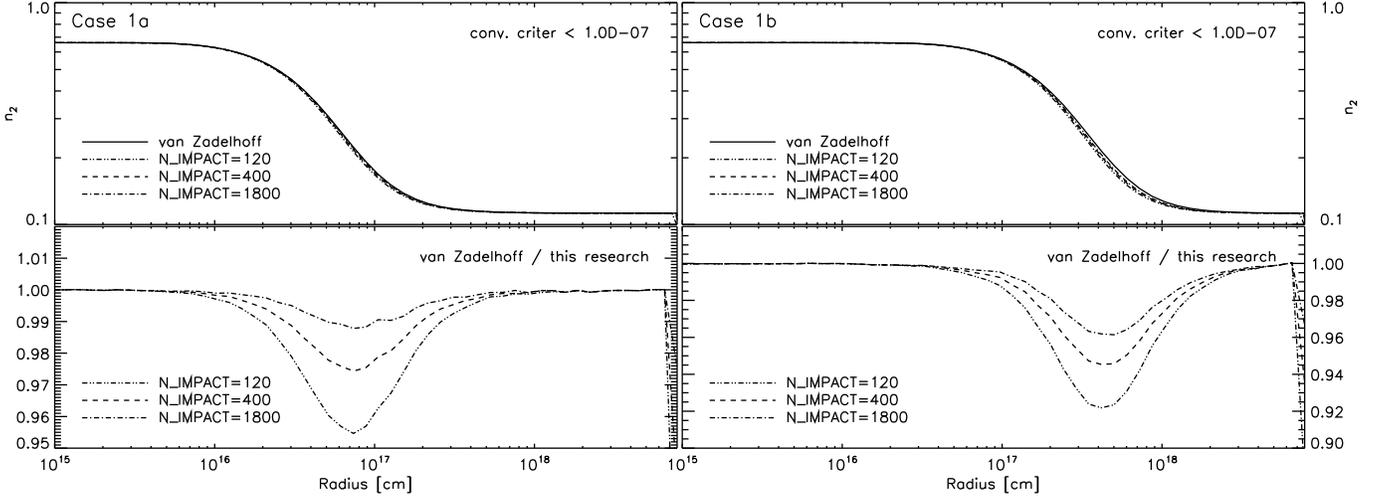}
  \caption{Comparison of the benchmark case 1a (left) and 1b (right)
    as defined in \citet{vanZadelhoff2002A&A...395..373V} for
    different spatial resolutions. {\underline{Upper panel:}} number
    density $n_2$ in the upper level of the fictive 2-level molecule
    as a function of radial distance from the star. The full line
    represents the mean of the number densities of all codes discussed
    by \citet{vanZadelhoff2002A&A...395..373V}. {\underline{Lower
        panel:}} ratio between their and our result. For increasing
    resolution the solutions converge. \emph{Dotted line:} $\Delta R/R
    \approx 0.07$; \emph{dashed line:} $\Delta R/R \approx 0.02$;
    \emph{dashed-dotted line} $\Delta R/R \approx 0.005$. }
  \label{benchmarkfig}
\end{figure*}

%%%%%%%%%%%%%%%%%%%%%%%%%%%%%%%%%%%%%%%%%%%%%%%%%%%%%%%%%%%%%%%%%%%%%%%%%%%
\subsection{The molecular model}
The rate coefficients for collisions of CO with H$_2$ at kinetic
temperatures from 10 to 4000\,K are taken from
\citet{Larsson2002A&A...386.1055L}. In this work, rotational quantum
numbers up to $J_u = 40$ were taken into account. Data on the
vibrational-rotational $v = 1 - 0$ and the rotational transitions are
from the HITRAN database \citep{Rothman1987ApOpt..26.4058R}. A total
of 62 levels and 120 transitions are treated, of which 31 levels are
in the ground state of $^{12}$C$^{16}$O, and 31 levels in the first
vibrational state. Comparison with the CO line list of
\citet{Goorvitch1994ApJS...91..483G} shows that the used Einstein
$A$-coefficients are in general good agreement; they are however
systematically lower than the ones of
\citet{Goorvitch1994ApJS...91..483G}, with a mean difference of a
factor of 1.017, a maximum of a factor 1.035, and a minimum of a
factor 1.005.
 
%%%%%%%%%%%%%%%%%%%%%%%%%%%%%%%%%%%%%%%%%%%%%%%%%%%%%%%%%%%%%%%%%%%%%%%%%%%
\subsection{Benchmarking}
To test the accuracy, the new code has been benchmarked against
existing radiative transfer codes. We used the study performed by
\citet{vanZadelhoff2002A&A...395..373V}, who compared a number of
independent computer programs for radiative transfer in molecular
rotational lines. None of the codes discussed in this comparison
solves the radiative transfer in the comoving frame. While in the rest
(inertial) frame the precision of a numerical code can be increased by
using a second-order (or higher) integration of the transfer equation,
in the comoving frame formalism it is written as two first-order
differential equations yielding only first-order accuracy
\citep{Mihalas1975ApJ...202..465M}. In the case study of problems
1a/1b [having a simple power law density, constant temperature and a
fictive 2-level molecule], the use of a first-order integration
introduces errors of the order of 12\,\%
\citep{vanZadelhoff2002A&A...395..373V}. Hence, for a proper
benchmarking of the code, the spatial resolution in the CMF should be
increased to `mimic' a higher-order solution, i.e. to improve the
accuracy. This is illustrated in Fig.~\ref{benchmarkfig}, for case 1a
and 1b. It is clear that higher spatial resolution results in a better
agreement with the other (non-CMF) codes, reaching a 1.2\,\% level of
accuracy for $\Delta R/R \approx 0.005$.

Case 2 of \citet{vanZadelhoff2002A&A...395..373V}, describing an
inside-out collapsing envelope observed in rotational transitions of
HCO$^+$, has not been simulated since the CMF method is designed for a
continuous \emph{in}creasing velocity, implying a blue-wing condition
stating that the highest frequency point in the local profile can
intercept only continuum radiation from other points in the
atmosphere.

%%%%%%%%%%%%%%%%%%%%%%%%%%%%%%%%%%%%%%%%%%%%%%%%%%%%%%%%%%%%%%%%%%%%%%%%%%%
\subsection{Effect of variable mass loss on the predicted line profiles of CO}
\label{var_mass2}
\begin{figure}[!thp]
  \includegraphics{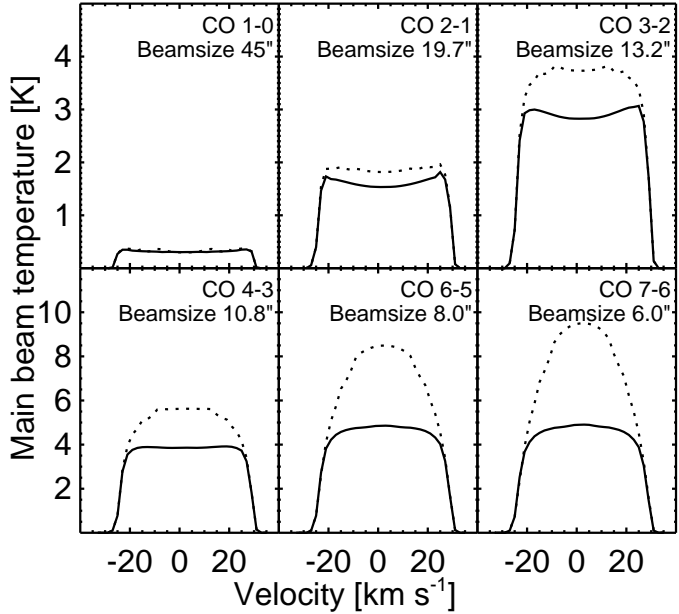}
  \caption{CO rotational line profiles (a) full line: for \object{GX Mon} using
    $T(r)$ and $v(r)$ of Fig.~\ref{graph_constant}, (b) dotted line: in
    case of a variable mass-loss rate as displayed in
    Fig.~\ref{graph_2step_high}. Note the different ordinate axes
    between upper and lower panels.}
  \label{profiles_2step_high}
\end{figure}

\begin{figure*}[htp]
  \includegraphics[height=\textwidth,angle=90]{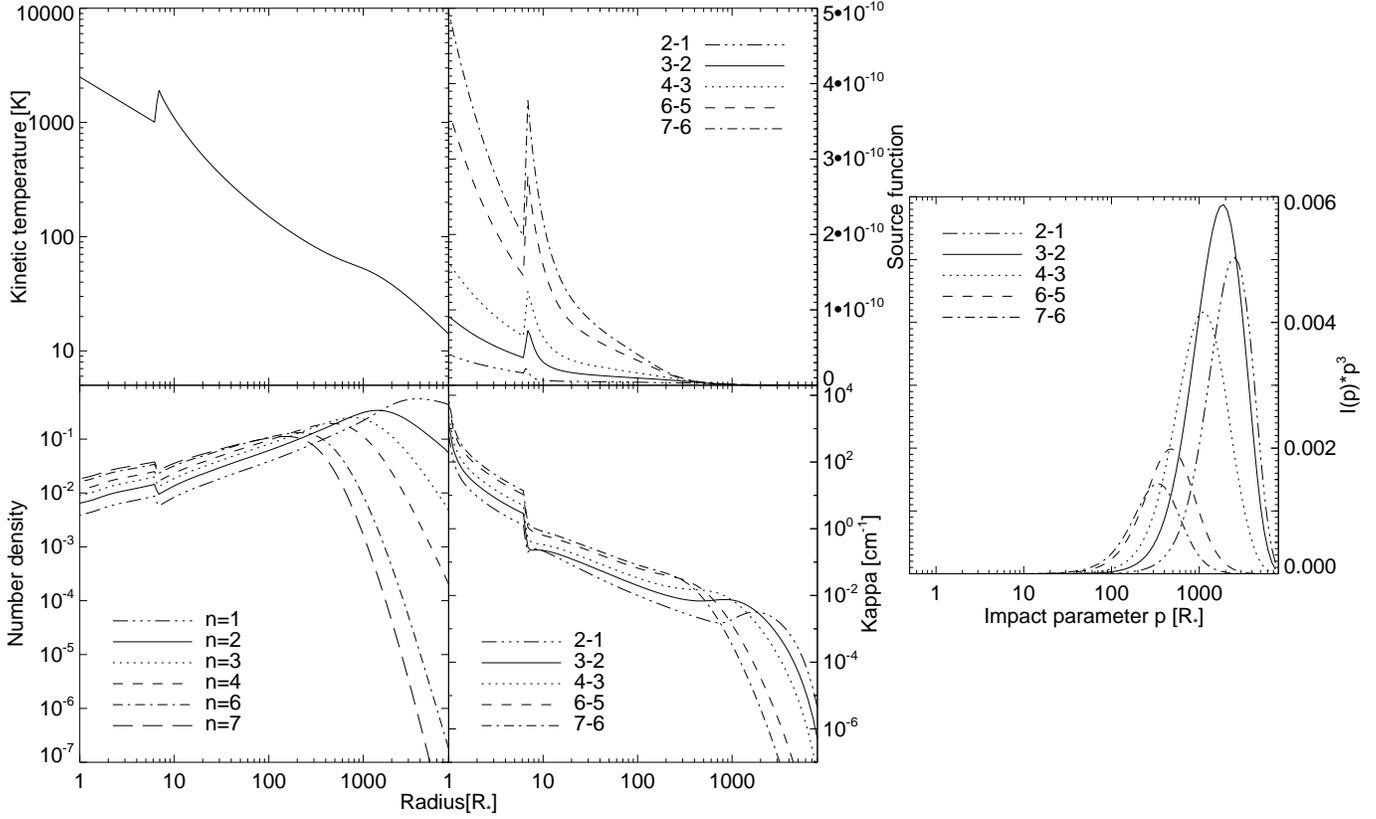}
  \caption{Detailed look at the wind structure of \object{GX Mon} for a
    constant mass loss of $7.2 \times 10^{-6}$\,\Msun\,yr$^{-1}$. Temperature
    [K], normalised number density [cm$^{-3}$], source function
    [erg\,s$^{-1}$\,cm$^{-2}$\,Hz$^{-1}$\,ster$^{-1}$], and the
    absorption coefficient $\kappa_{\nu}$ [cm$^{-1}$] as a function of
    radius; $I(p)*p^3$ as function of the impact parameter $p$ ---
    using $T(r)$ and $v(r)$ of Fig.~\ref{graph_constant}.}
  \label{line_contrib_constant}
\end{figure*}

\begin{figure*}[!thp]
  \includegraphics[height=\textwidth,angle=90]{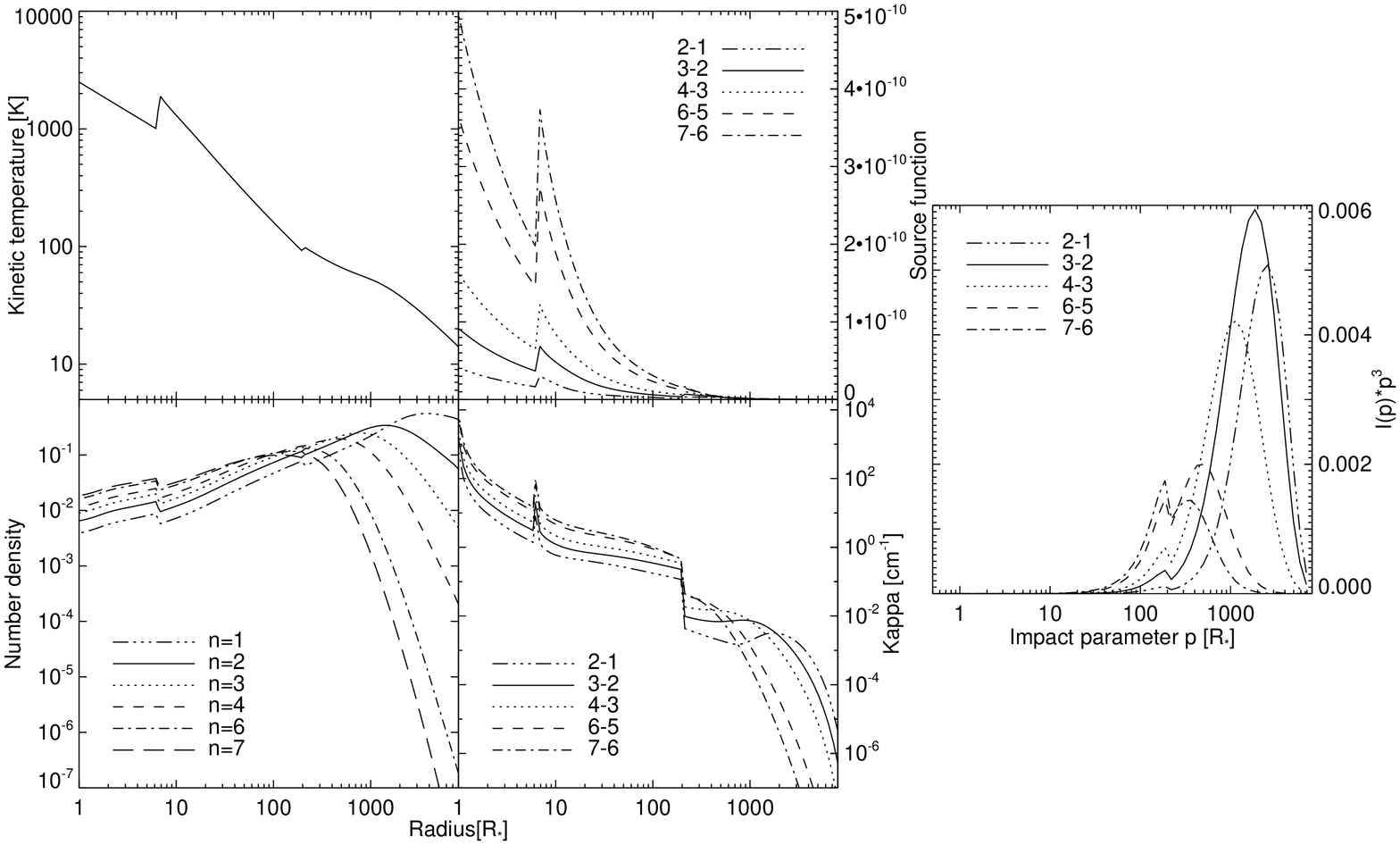}
  \caption{\label{line_contrib_2step_high} Detailed look at the wind
    structure of a superwind phase for \object{GX Mon}. The superwind
    increases the mass loss by a factor 10 from the stellar surface
    up to 200\,\Rstar. Temperature [K], normalised number density
    [cm$^{-3}$], source function $S_{\nu}$
    [erg\,s$^{-1}$\,cm$^{-2}$\,Hz$^{-1}$\,ster$^{-1}$], and the
    absorption coefficient $\kappa_{\nu}$ [cm$^{-1}$] as a function
    of radius; $I(p)*p^3$ as function of the impact parameter $p$.}
\end{figure*}

For the specific case of \object{GX Mon} and the simulation of a superwind
phase with the same parameters as in Sect.~\ref{var_mass1} the
calculated low excitation CO line profiles are displayed in
Fig.~\ref{profiles_2step_high}. The adopted beam sizes are those for
the JCMT, and are indicated in the upper corner of each panel. Number
densities, source functions $S_{\nu}$ and absorption coefficient
$\kappa_{\nu}$ as a function of radius are plotted in
Figs.~\ref{line_contrib_constant} and \ref{line_contrib_2step_high}.

A superwind phase increases the number densities of all the energy
levels from $R_{\rm{inner}}$ to 200\,\Rstar, where the high mass-loss
steps.

The sudden rise in temperature at the start of dust condensation gives
a strong peak in the source function $S_{\nu}$ at R$_{\rm{inner}}$.
The higher temperature in the superwind phase, results in (1) higher
values of the source function up to $200$\,\Rstar, and lower values
thereafter, and (2) a strong increase in the opacity up to
200\,\Rstar\ due to the higher number densities till the sudden
decrease in mass loss rate introduces a sharp decrease in opacity.

The figures displaying $I(p)*p^3$ are instructive plots to illustrate
where the lines mostly originate \citep[see
also][]{Kemper2003A&A...407..609K}. In case of a constant mass-loss
rate, there is only one peak in $I(p)*p^3$, being closer to the star
for higher excitation lines; for the superwind case, we see a second
peak arising around 200\,\Rstar, the end of the enhanced mass loss. A
direct result from the enhancement in number densities at distances
$\le 200$\,\Rstar\ is an increase in integrated intensity for the
superwind case, especially for the higher excitational rotational
lines which are formed closer in. Moreover, the sudden change in
opacity (number density) at 200\,\Rstar\ gives rise to `bumps' in the
line profiles, here with (projected) velocities around 10 and
15\,\kms.

Depending on the stellar input parameters, the abruptness of the
variation in \Mdot, and the place where (or time when) this change has
taken place different kind of `composite' profiles (i.e.\ displaying
``dips and bumps'' in the line profile) are seen. This kind of
composite profiles is also seen in observational data, with \object{$\alpha$
Ori} and \object{VY CMa} being examples in \citet{Kemper2003A&A...407..609K},
and other targets in \citet{Winters2003A&A...409..715W}. Note that the
relative strength of the observed profiles of some stars, where this
kind of `composite' profile is not visible, is also indicative for a
variability in the mass-loss rate \citep[see the discussion on \object{WX Psc}
in][]{Kemper2003A&A...407..609K}.

The complex interplay between stellar and mass-loss parameters and
some general trends linking the mass-loss rate to the observed line
intensity will be described in a forthcoming paper.

%%%%%%%%%%%%%%%%%%%%%%%%%%%%%%%%%%%%%%%%%%%%%%%%%%%%%%%%%%%%%%%%%%%%%%%%%%%
\section{\object{VY CMa}}\label{vycma}
To illustrate the diagnostic strength of the rotational CO profiles,
we now apply our code to model the observed CO rotational line
transitions emitted in the CSE around \object{VY CMa}. In
Sect.~\ref{basicparVYCMa}, the basic parameters of this supergiant are
summarised. The CO line profiles used in the analysis are described in
Sect.~\ref{obserVYCMa}. Sections \ref{spherVYCMa} and \ref{modVYCMa}
are devoted to the modelling of the CO profiles. We will demonstrate
that it is possible to derive the mass-loss \emph{history} of this
target in great detail. Our results are compared to other studies in
Sect.~\ref{compVYCMa}, and discussed in Sect.~\ref{discVYCMa}.

%%%%%%%%%%%%%%%%%%%%%%%%%%%%%%%%%%%%%%%%%%%%%%%%%%%%%%%%%%%%%%%%%%%%%%%%%%%
\subsection{Basic parameters of \object{VY CMa}}
\label{basicparVYCMa}
\begin{table}
  \caption{\label{param_VYCMa} Parameters of \object{VY CMa}. We have
    assumed cosmic carbon and oxygen abundances
    \citep{Anders1989GeCoA..53..197A}. } 
  \begin{tabular}{lll}\hline \hline
    \rule[-3mm]{0mm}{8mm}Parameter & This work & \citet{Smith2001AJ....121.1111S}\\ 
    \hline
    \Tstar & 2800\,K  & 3000\,K \\
    \Rstar & $1.6 \times 10^{14}$\,cm & $2 \times 10^{13}$\,cm \\ 
    \Lstar & $3 \times 10^{5}$\,\Lsun & $1.6 \times 10^{5}$\,\Lsun \\ 
    \Mstar & 15\,\Msun & 15\,\Msun \\
    R$_{\rm{inner}}$  & 10\,\Rstar & 7\,\Rstar \\
    \Mdot & see
    Fig.~\ref{VYCMa_mdot} & $10^{-5} \times 10^{-4}$\,M$_{\odot}$\,yr$^{-1}$ till $0.5''$ \\
    \Mdot & see
    Fig.~\ref{VYCMa_mdot} & $3 \times 10^{-4}$\,M$_{\odot}$\,yr$^{-1}$ from $0.5''$ up to $7''$ \\
    v$_{\rm{stoch}}$ & 1\,\kms & 1\,\kms \\
    v$_{\infty}$ & 35\,km\,s$^{-1}$ & 35\,km\,s$^{-1}$ \\
    distance & 1500\,pc & 1500\,pc \\
    \rule[-3mm]{0mm}{3mm}$\psi$ & 0.002 & 0.01 \\
    \hline
  \end{tabular}
\end{table}

\object{VY CMa} is a highly luminous, variable M2/3II supergiant. The star is
heavily obscured, with only 1 percent of the total luminosity being
detected at optical wavelengths \citep{LeSidaner1996A&A...314..896L}.
The distance, luminosity, effective temperature and stellar radius
have been estimated by several groups.
These studies seem to converge on a distance of 1500\,pc, a luminosity
between $2 - 5 \times 10^5$\,\Lsun, and an effective temperature of
2800\,K \citep{Harwit2001ApJ...557..844H,
  LeSidaner1996A&A...314..896L, Monnier1999ApJ...512..351M}. The
current mass of \object{VY CMa} is estimated to be 15\,\Msun\
\citep{Wittkowski1998A&A...340L..39W}. These values are adopted for
the stellar parameters of \object{VY CMa}; they are summarised in Col.~2 in
Table~\ref{param_VYCMa}.
 
The OH, H$_2$O, and SiO maser emission indicate expansion velocities
between $\sim 30$ and $\sim 40$\,\kms \citep{Reid1976ApJ...209..505R,
  Snyder1975ApJ...197..329S, Reid1978ApJ...220..229R}. Modelling of CO
rotational lines, OH maser emission and (near)-infrared data tracing
the dust properties indicate values for the mass loss between $2.3
\times 10^{-5}$ and $4 \times 10^{-4}$\,\Msun\,yr$^{-1}$ (see
Table~\ref{masslossrates}). \object{VY CMa}'s very high mass-loss rate fuels
its optically thick circumstellar envelope and creates a rich infrared
excess spectrum \citep[see the 2.4 -- 200\,$\mu$m ISO spectrum in
][]{Harwit2001ApJ...557..844H}: metallic Fe and amorphous silicates
dominate the infrared spectrum, with possible minor contributions from
crystalline forsterite and crystalline enstatite.

\begin{table*}[!thp]
  \caption{\label{masslossrates} Mass-loss rates for \object{VY CMa} as found in
    literature and scaled to a distance of 1500\,pc. }
  {\footnotesize{
      \begin{tabular}{l|llll}\hline \hline
        \rule[-3mm]{0mm}{8mm}& Observ. data & \Mdot\ [\Msun\,yr$^{-1}$] &
        comments & Reference  \\ 
        \hline 
        & OH maser emission & $1.2 \times 10^{-4}$ & from OH shell radius +
        $n_{\rm{H}} = 10^4$\,cm$^{-3}$ & \citet{Booth1984IAUS..110..313B}\\
        & CO (2--1) & $1\times
        10^{-4}$ & scaling based on analysis of &
        \citet{Zuckerman1986ApJ...304..394Z} \\ 
        \raisebox{-1.ex}[0pt]{\rotatebox{90}{gas}}& & & \citet{{Knapp1985ApJ...292..640K}} & \\
        & CO(1--0) & $2.3 \times 10^{-5}$ & from analytical expression &
        \citet{Loup1993AAS...99..291L} \\
        \hline
        & spectrosc.\ data 2 -- 2.5\,\mic\ +&  $2.3 \times 10^{-4}$ & data from
        \citet{Hyland1972AandA....16..204H}; &
        \citet{Bowers1983ApJ...274..733B}\\ 
        & photometr.\ data till 20\,\mic & & $\psi = 0.01$  & \\
        & submm data at 400\,\mic & & \Mdot$_{\rm{dust}} = 7.8 \times
        10^{-7}$\,\Msun\,yr$^{-1}$ & \citet{Sopka1989AA...210...78S} \\
        \raisebox{0.ex}[0pt]{\rotatebox{90}{dust}} & ISI 11\,\mic\ interf. & $3.1 \times 10^{-4}$  & $\psi = 0.005$ &
        \citet{Danchi1994AJ....107.1469D} \\
        & spectrosc.\ data 2.38 -- 200\,\mic & $4 \times 10^{-4}$ & $\psi
        = 0.01$ & \citet{Harwit2001ApJ...557..844H} \\
        & surface photometry & $10^{-5} - 10^{-4}$ & till
        $0.5''$ & \citet{Smith2001AJ....121.1111S} \\
        & & $3 \times 10^{-4}$ & from $0.5''$ up to $7''$\ \ \ \raisebox{1.5ex}[0pt]{$\psi=0.01$} & \\
        \hline
      \end{tabular}
    }}
\end{table*}

%%%%%%%%%%%%%%%%%%%%%%%%%%%%%%%%%%%%%%%%%%%%%%%%%%%%%%%%%%%%%%%%%%%%%%%%%%%
\subsection{Observational data of \object{VY CMa}}
\label{obserVYCMa}
The rotational CO (2--1), (3--2), (6--5) and (7--6) line profiles have
been observed by \citet{Kemper2003A&A...407..609K} with the JCMT and
the CO (1--0) line by \citet{Nyman1992AAS...93..121N} with the SEST.
The flux calibration accuracy for the JCMT data was estimated to be
10\,\% for the CO (2--1) and (3--2) lines, and $\sim 30\,\%$ for the
CO (6--5) and (7--6) lines since no reliable flux standards are
available for these higher excitation lines. The CO(1--0) line has a
main beam temperature of 0.06\,K, with an rms noise of 0.019\,K
\citep{Nyman1992AAS...93..121N}. As can be seen from
Fig.~\ref{VYCMa_model}, the profiles show a complex structure.
Especially the different peaks present in the CO (2--1) and (3--2)
profiles are indicative for \emph{variation(s) in the mass-loss rate}.

%%%%%%%%%%%%%%%%%%%%%%%%%%%%%%%%%%%%%%%%%%%%%%%%%%%%%%%%%%%%%%%%%%%%%%%%%%%
\subsection{A spherically symmetric envelope?}
\label{spherVYCMa}
There exist indications that the CSE surrounding \object{VY CMa} is not
spherically symmetric. While some authors interpret the complex
optical and infrared data in terms of an expanding disk or a bipolar
outflow near the central object \citep[$< 0.2''$;
e.g.][]{Herbig1972ApJ...172..375H, Bowers1983ApJ...274..733B,
  Efstathiou1990MNRAS.245..275E}, other data do not reveal indications
for a disk-like or bipolar geometry
\citep[e.g.][]{Monnier1999ApJ...512..351M}. Recently, \emph{``Hubble
  Space Telescope''} in combination with ground-based infrared images
revealed a complex structure of knots and filamentary arcs in the
asymmetric reflection nebula several arcseconds across, around the
obscured central star \citep{Smith2001AJ....121.1111S}.

Nevertheless, we will assume spherical symmetry in modelling the
molecular lines arising from the CSE. This is justified with the
following arguments: (1) If our assumption of spherical symmetry is
incorrect, this is more so for the innermost regions, where there are
indications of a disk with extent $\la 0.2''$ ($\sim 30$\,\Rstar). We
are however aiming at a reconstruction of the mass-loss history from
low-excitation CO lines, which are formed further out, and trace a
region up to a few thousand stellar radii. (2) If the inhomogeneities
are due to blobs of which the dimensions are modest compared to the
total extent of the wind and which are more or less homogeneously
distributed in the wind, the use of a spherically symmetric model
gives us the average of the mass loss over all directions. In that
way, our spherical model gives us an idea on the general mass-loss
history of \object{VY CMa}. (3) Since the beam profiles of the instruments used
to observe the CO lines are quite large (with a half power beam width
(HPBW) of $45''$ in the CO(1--0) line, $19.7''$ in the CO(2--1),
$13.2''$ in the CO(3--2), $8.0''$ in the CO(6--5) and $6.0''$ in the
CO(7--6) line) the CO line profiles can not be used to trace
small-scale structure, but only to estimate the density, temperature
and velocity structure averaged over all directions in the regions
where they are formed.

Moreover, \citet{Smith2001AJ....121.1111S} demonstrated that the
\emph{apparent} large-scale asymmetry of the nebula in the optical and
near-IR images results from a combination of high extinction and
backscattering by dust grains. Thermal-infrared images ($5 -
10$\,$\mu$m) reveal a more symmetric distribution.

%%%%%%%%%%%%%%%%%%%%%%%%%%%%%%%%%%%%%%%%%%%%%%%%%%%%%%%%%%%%%%%%%%%%%%%%%%%
\subsection{Modelling the rotational CO line profiles}\label{modVYCMa}
\begin{figure}[!thp]
  \includegraphics[width=8.8cm]{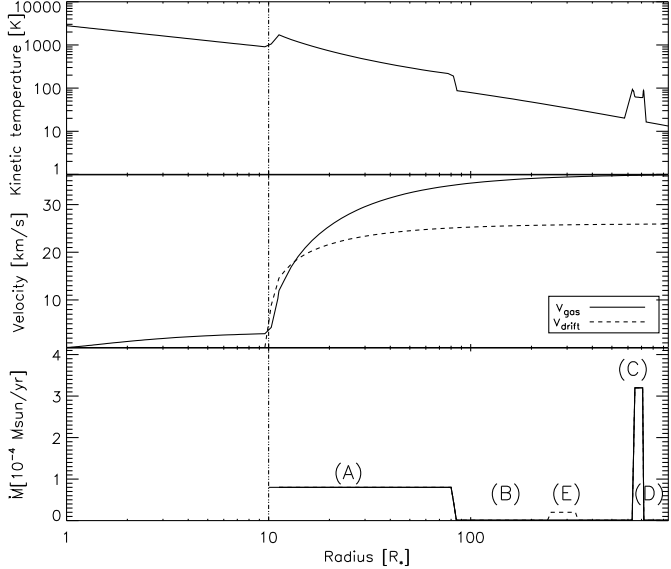}
  \caption{\emph{Upper:} Estimated temperature profile, \emph{middle:}
    estimated velocity structure, and \emph{bottom:} estimated
    mass-loss rate \Mdot$(r)$ for \object{VY CMa} as a function of radial
    distance to the star (black line). The blue/grey line in the panel
    displaying the mass-loss history represents a model with an extra
    enhancement in mass loss (\Mdot$(r) = 2 \times 10^{-5}$\,\Msun\,yr$^{-1}$)
    around 300\,\Rstar\ having the same theoretical line profiles as
    in Fig.~\ref{VYCMa_model}.}
  \label{VYCMa_mdot}
\end{figure}
\begin{figure*}[!thp]
  \sidecaption
  \includegraphics[height=12cm,angle=90]{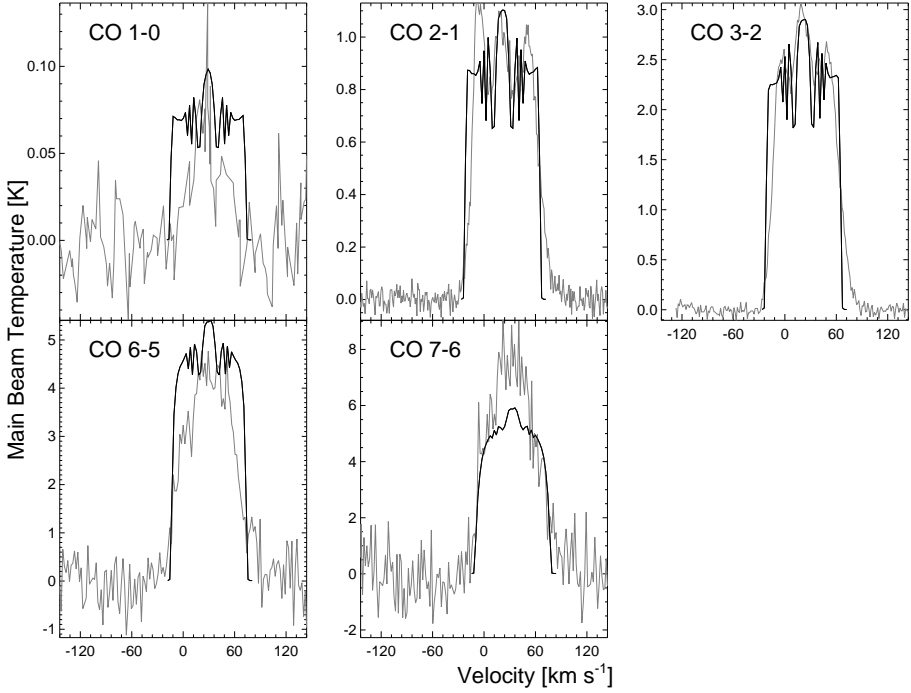}
  \caption{CO rotational line profiles of \object{VY CMa}
    \citep{Kemper2003A&A...407..609K} (grey line) compared
    with the model predictions (black line) based on the parameters
    of our best model (see Table~\ref{param_VYCMa}). }
  \label{VYCMa_model} 
\end{figure*}
To determine the mass-loss history of \object{VY CMa} a first initial grid of
$\sim 20000$ models, with stellar parameters as specified in Col.~2 in
Table~\ref{param_VYCMa}, was run to find the global minima in
parameter-space using a classical reduced $\chi^2$-method. 
In this, we treat the integrated strength of the line and the line
shape as two separate entities to be fitted. This allows to give more
weight to the line profile which is more reliable than the absolute
flux calibration. The mass-loss parameters of the
best models were then further refined.

Parameters characterising the CSE of the best model are given in the
second column of Table~\ref{param_VYCMa}. Since the calibration
uncertainty on the CO(2--1) and (3--2) line is smaller than for the
other profiles, highest weight is given automatically to these lines
in the reduced $\chi^2$-test. The derived temperature profile,
velocity structure and mass-loss history are displayed in
Fig.~\ref{VYCMa_mdot}; a comparison between observed and theoretical
line profiles is shown in Fig.~\ref{VYCMa_model}.

Before comparing to results of other studies, we first discuss the
derived \Mdot-profile and the associated uncertainties. The
\Mdot-profile has a value of $8 \times 10^{-5}$\,\Msun\,yr$^{-1}$ till
80\,\Rstar\ (denoted as (A) in Fig.~\ref{VYCMa_mdot}). Mass-loss rates
in the order of $\sim 1 \times 10^{-4}$\,\Msun\,yr$^{-1}$ in these inner
regions still result in acceptable line profiles for the CO(6--5) and
(7--6) line. Between $\sim 80$ and $\sim 600$\,\Rstar, we determine a
low mass-loss rate, being of the order of $1 - 3 \times
10^{-6}$\,\Msun\,yr$^{-1}$ (denoted as (B) in Fig.~\ref{VYCMa_mdot}). This
mass-loss value can not be below $5 \times 10^{-7}$\,\Msun\,yr$^{-1}$, since
the total dust content at the base of the wind would be too low to
drive the stellar wind. This period of low mass-loss rate was preceded
(in time) by a short period (between 80 and 100\,yr at a expansion
velocity of 35\,\kms) of very strong mass loss in the order of $3 -
3.5 \times 10^{-4}$\,\Msun\,yr$^{-1}$ (region (C)). This period is situated
around 680\,\Rstar\ ($\sim 1000$\,yr ago). One can simulate almost
analogous line profiles with a somewhat smaller \Mdot\ (around $2.5
\times 10^{-4}$\,\Msun\,yr$^{-1}$) having lasted somewhat longer ($\sim
150$\,yr). This period of strong mass loss can however not be placed
closer to the star, since the predicted CO(6--5) line becomes far too
bright. The estimated mass-loss rate in region (D) is $\sim 1 \times
10^{-6}$\,\Msun\,yr$^{-1}$.

\begin{figure*}[!thp]
  \sidecaption
  \includegraphics[height=12cm,angle=90]{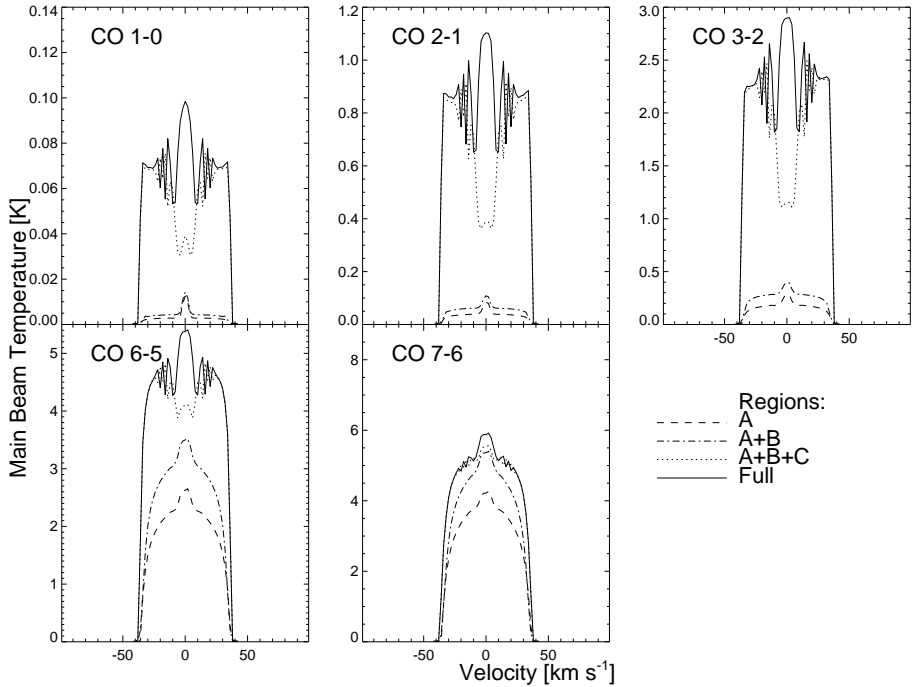}
  \caption{CO rotational line profiles resulting from the best model
    (full line) compared to theoretical CO line profiles taking into
    account (1) region (A): dashed line, (2) regions (A)+(B):
    dashed-dotted line, and (3) regions (A)+(B)+(C): dotted line. }
  \label{VYCMa_decompose} 
\end{figure*}

The estimated kinetic temperature, velocity and mass-loss profile
yield the predicted CO line profiles displayed in
Fig.~\ref{VYCMa_model}. The central peak (with a width of $\sim
20$\,km\,s$^{-1}$) visible in all line profiles originates close to the star
($r < 15\,$\Rstar), where the dust is accelerated away due to
radiation pressure. Region (A) is responsible for the main
contribution to the high-excitation CO(7--6) and (6--5) lines. With a
temperature above $\sim 100$\,K, the density in this region results in
an optically-thick, non-resolved Gaussian line profile (see dashed
line in Fig.~\ref{VYCMa_decompose}). The influence of region (B) is
only marginal, and results in a small intensity enhancement in all
line profiles (see dashed-dotted line in Fig.~\ref{VYCMa_decompose}).
The region with a high density enhancement, denoted as region (C), and
with an extent from $\sim 600$ to $\sim 700$\,\Rstar, contributes
differently to the high-excitation CO(7--6) and (6--5) lines than to
the lower-excitation lines. In the former case, the temperature ($<
90$\,K) in region (C) is too low to significantly populate the
higher-excitation energy levels, yielding an optically thin, resolved
double-peaked line profile, for the beamwidths of the JCMT. The
situation is markedly different for the lower-excitation lines: (1)
the low temperature in this region causes $\sim 5\,\%$ of the total CO
number density to be in the ground state, $\sim 13\,\%$ to be in the
$J=1$ state, $\sim 21\,\%$ in the $J=2$ state and $\sim 23\,\%$ in the
$J=3$ state, yielding optical depths larger than 1. The shape of the
line profiles then follows the shape of the source function $S_{\nu}$,
which form can be directly inferred from the temperature distribution
displayed in the upper panel of Fig.~\ref{VYCMa_mdot}. The increase
and subsequent decrease in $T(r)$ in region (C) results in a
double-peaked line profile, with widths in the red and blue peak of
some 20\,km\,s$^{-1}$. As an illustration, we have plotted the contribution of
region (A)+(B)+(C) to the full line profile in
Fig.~\ref{VYCMa_decompose}. Note that only the (3--2) line transition
is slightly resolved in region (C), while the larger beamwidths for
the (2--1) and (1--0) line do not resolve the envelope. It can be seen
in Fig.~\ref{VYCMa_decompose} that the intensity in the central peak
is under-predicted in the CO(2--1) and (3--2) line (dotted line in
Fig.~\ref{VYCMa_decompose}). The low quality of the CO(1--0) line
makes it difficult to substantiate this for this line. The
contribution from region (D) fills in this gap, due to the still
substantial mass-loss rate of \Mdot$(r) = 1 \times 10^{-6}$\,\Mdot\,yr$^{-1}$
in this region and its low temperature.

The small scale ripples visible in the calculated line profiles are
not due to numerical noise in the radiative transfer and can not be
removed by increasing the number of grid-points (now taken to 150
gridpoints). The cause for the
presence of these ripples is the complex temperature structure in
region (C) which results from the competing effects of heating and
cooling processes. The associated source function is therefore quite
complex. In combination with small scale structure in the opacity
resulting from the number-density profile this explains these ripples
in the theoretical profiles. However, a less abrupt \Mdot-profile
would cause these ripples to decrease in amplitude.

The match between observations and theoretical predictions is not
perfect. Main reasons for this are the simplifying assumption of a
spherical symmetric envelope and a velocity structure being
continuously increasing with $r$, as required by the comoving frame
formalism. In reality the higher mass-loss rate in region (C) should
drive a dusty wind with a different velocity than in the low-mass
regions and maybe also with another dust-to-gas ratio.

\subsection{Comparison with other studies}\label{compVYCMa}
A direct comparison with other derived mass-loss rates (see
Table~\ref{masslossrates}) is difficult, since all of the studies
mentioned in Table~\ref{masslossrates} assume a density varying as
$r^{-2}$. In case of the analysis of dust spectra, one moreover has to
assume a dust-to-gas ratio $\psi$ to estimate the gas mass-loss rate.
The main differences with our study are that in our case (1) the
mass-loss rate is variable, and hence the density is not varying as
$r^{-2}$, (2) the dust-to-gas ratio is derived self-consistently to
obtain the observed expansion velocity, and (3) the IR emission traces
warm dust close to the star, while the analysis of different
low-excitation rotational CO lines gives us diagnostics on a much
broader spatial region.

\begin{figure*}[!thp]
  \sidecaption
  \includegraphics[height=12cm,angle=90]{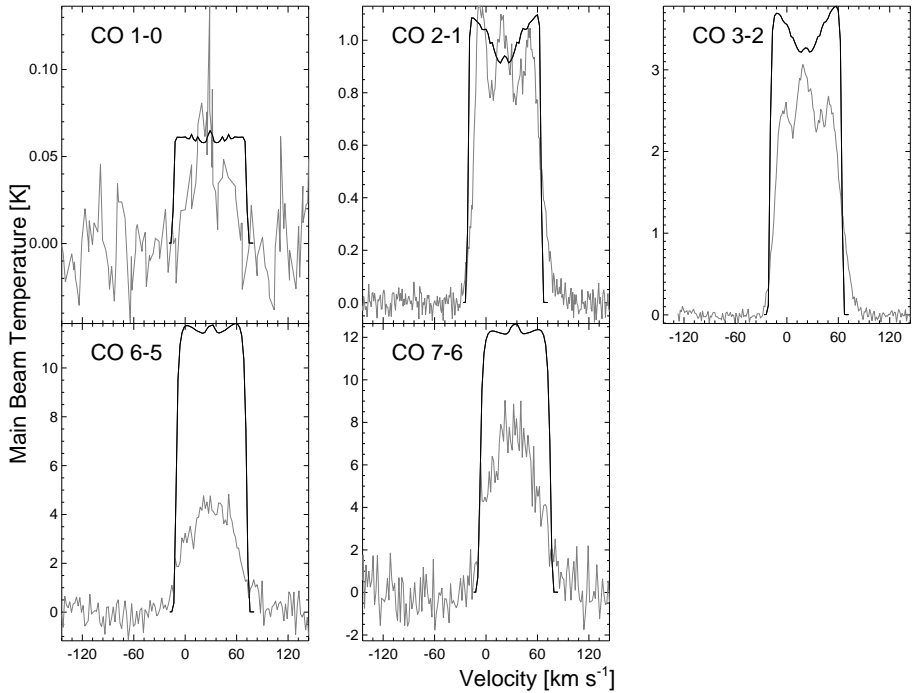}
  \caption{CO rotational line profiles of \object{VY CMa}
    \citep{Kemper2003A&A...407..609K, Nyman1992AAS...93..121N}
    (grey line) compared with the model predictions (black line)
    based on the stellar parameters of \object{VY CMa} as given in the second
    column of Table~\ref{param_VYCMa}, but with a constant \Mdot\ of
    $3.3 \times 10^{-5}$\,\Msun\,yr$^{-1}$ (black line). The derived
    dust-to-gas ratio $\psi$ is 0.002.}
  \label{VYCMa_model_constant} 
\end{figure*}

Assuming a constant mass-loss rate, the best fit model is obtained at
\Mdot\,=\,$3.3 \times 10^{-5}$\,\Msun\,yr$^{-1}$ (see
Fig.~\ref{VYCMa_model_constant}). From the discrepancy in strength of
the higher-excitation rotational CO line transitions, it is
immediately clear that a model with constant mass loss is unable to
explain the observed lines. Moreover, this model fails completely in
fitting the detailed line profiles. This is not surprising in view of
what is already known about this object (see Sect.~\ref{spherVYCMa}).
This value for \Mdot\ can be directly compared to the result of
\citet{Loup1993AAS...99..291L}, who derived a value of $2.3 \times
10^{-5}$\,\Msun\,yr$^{-1}$ (scaled to a distance of 1500\,pc) from the
CO(1--0) integrated line intensity.
\citet{Zuckerman1986ApJ...304..394Z} used the CO(2--1) integrated line
intensity as observed with the NRAO 12\,m Kitt Peak to derive a much
higher mass-loss rate of $\sim 10^{-4}$\,\Msun\,yr$^{-1}$. The noise in their
observed line profiles was so large that the clear composite nature as
seen in Fig.~\ref{VYCMa_model} is not traced. As
\citet{Zuckerman1986ApJ...304..394Z} stated, they scaled the analysis
of \citet{Knapp1985ApJ...292..640K} for other stars that are located
at comparable distances from Earth ($\sim 1500$\,pc). However,
\citet{Knapp1985ApJ...292..640K} derived mass-loss rates from CO(1--0)
lines of a sample of 50 stars, of which only 2 were supergiants. A key
point resulting from our study is that the analysis of \emph{full line
  profiles} with different excitation energies provide a much better
diagnostics to derive the mass-loss rate than peak intensities, also
taking into account that the main calibration uncertainties do not
effect the line shape.

Values for the mass-loss rate traced by dust characteristics are
generally larger than our mean mass-loss rate of $5.1 \times
10^{-5}$\,\Msun\,yr$^{-1}$, especially keeping in mind that the dust-to-gas
ratio used by other authors (see Table~\ref{masslossrates}) is always
higher than the value we infer from the CO-lines ($\psi = 0.002$).
However, (near)-infrared dust emission traces regions closer to the
star than the low-excitation rotational CO lines. The mean mass loss
rate up to 100\,\Rstar\ that we determine is indeed about $0.75 (\pm
0.25) \times 10^{-4}$\,\Msun\,yr$^{-1}$, substantially higher than the mean
mass-loss rate averaged over the full envelope of $5.1 \times
10^{-5}$\,\Msun\,yr$^{-1}$.

%%%%%%%%%%%%%%%%%%%%%%%%%%%%%%%%%%%%%%%%%%%%%%%%%%%%%%%%%%%%%%%%%%%%

\subsection{Discussion: mass-loss variability}\label{discVYCMa}
\citet{Smith2001AJ....121.1111S} also studied mass-loss variability.
From surface photometry they derived that the star may have
experienced enhanced mass loss over the past 1000\,yr ($\sim 3 \times
10^{-4}$\,\Msun\,yr$^{-1}$ between $\sim 0.5''$ and $\sim 7''$), with
indications of a lower mass loss (in the order of $10^{-5}$ to
$10^{-4}$\,\Msun\,yr$^{-1}$ till $\sim 0.5''$) during the past 100\,yr (see
third column in Table~\ref{param_VYCMa}). They noted that these
estimates can be off by a factor of a few due to uncertainties in the
calibration of the mid-IR imaging data, model assumptions and assumed
dust properties. Linking our derived mass-loss history with their
optical and near infrared images, the enhanced gas density traced in
our region (A) coincides with several bright nebular condensations
located at $1''$ to $2''$ south and southwest of the star (indicated
by `S' and `SW' in their Fig.~8). The high mass-loss phase indicated
by (C) in the bottom panel in Fig.~\ref{VYCMa_mdot} is situated at the
same radial distance as their `arc \#1', being from $5''$ to $7''$
from the star. The two other arcs showing up in their Fig.~8 (named as
`arc \#2' and `curved nebulous tail') at $2''$ to $4''$ from the star,
are not directly traced in our mass-loss history. However, test
computations show that an extra enhancement of mass loss around $3''$
with $\Mdot \le 2 \times 10^{-5}$\,\Msun\,yr$^{-1}$ with an extent of $1''$
may be present without seriously altering the calculated line profiles
(see dashed line in Fig.~\ref{VYCMa_mdot}, indicated by (E)). Hence,
we may conclude that our results are consistent with the Hubble and
near-IR images presented \citet{Smith2001AJ....121.1111S}.

We conclude from our analysis and the study performed by
\citet{Smith2001AJ....121.1111S}, that \object{VY CMa} underwent large
changes in mass-loss rate in the order of a factor 100 --- averaged
over all directions --- with a time separation of some 800\,yr. This
timeline is of the same order of magnitude as the separations of 200
-- 800\,yr observed by \citet{Mauron1999A&A...349..203M} observed in
the carbon-rich AGB-star \object{IRC +10 216}. Also
\citet{Schoier2005A&A...436..633S} found that the mass-loss rate
producing the faster moving wind in some carbon stars with detached
shells should be almost two orders of magnitude higher than the slower
AGB wind preceding this violent event. It is important to remark that
not all dust-driven winds do display this kind of modulations: a
sample of nine carbon-rich AGB-stars with mass-loss larger than $1.5
\times 10^{-5}$\,\Msun\,yr$^{-1}$ analysed by
\citet{Schoier2002A&A...391..577S} are not likely to have experienced
any drastic long-term mass-loss rate modulations over the past
thousands of years. As stated already in the introduction, this
modulation time scale for AGB stars is too large to be due to stellar
pulsations, and too short to be due to nuclear thermal pulses. It is
suggested by \citet{Simis2001A&A...371..205S} that such shells are
formed by a hydrodynamical oscillation in the envelope, while the star
is on the AGB. The cause of this effect is a subtle mechanism,
involving an intricate nonlinear interplay between gas-grain drift,
grain nucleation, radiation pressure, and envelope hydrodynamics.
Another mechanism proposed by \citet{Soker2000ApJ...540..436S} is a
solar-like magnetic activity cycle. The periodic change of direction
of the stellar magnetic field enables periodic structure formation,
with time intervals between consecutive ejection events in the order
of 200 -- 1000\,yr. In combination with giant convection cells which
may be present in the outer atmospheric layers of red supergiants
\citep{Freytag2002AN....323..213F}, this may explain the morphology of
the arcs \citep{Smith2001AJ....121.1111S}.

%%%%%%%%%%%%%%%%%%%%%%%%%%%%%%%%%%%%%%%%%%%%%%%%%%%%%%%%%%%%%%%%%%%%%%%%%%%
\section{Summary}
\label{summary}
We have developed a code to derive the gas mass-loss \emph{history}
from a non-LTE analysis of the full line profiles of CO rotational
transitions with different excitation energies, including a
self-consistent computation of the velocity structure and gas kinetic
temperature, taking several heating and cooling mechanisms in the CSE
into account. The code has been benchmarked to other existing codes,
demonstrating good performance.

The code is applied to the analysis of the wind structure of \object{VY CMa}.
For the first time, we could pinpoint the mass-loss \emph{history} of
this supergiant from the analysis of the CO(1--0), (2--1), (3--2),
(6--5), and (7--6) line profiles and could demonstrate that the
mass-loss rate underwent substantial changes during the last 1400\,yr.
We especially note a very strong peak of mass loss of $\sim 3.2 \times
10^{-4}$\,\Msun\,yr$^{-1}$ some 1000\,yr ago having lasted $\sim 100$\,yr.
This period of high mass loss was preceded by a long period of low
mass loss, in the order of $1 \times 10^{-6}$\,\Msun\,yr$^{-1}$ taking some
800\,yr. The current mass loss is estimated to be in the order of $5
\times 10^{-5}$ to $1 \times 10^{-4}$\,\Msun\,yr$^{-1}$. The mass-loss history
that we derive is significantly different from what one would estimate
from a constant mass-loss rate model, and therefore paints a much more
complete picture on the evolution of the mass loss of AGB and red
supergiant stars.

\begin{acknowledgements}
  LD acknowledges financial support from the Fund for Scientific
  Research - Flanders (Belgium), SH acknowledges financial support
  from the Interuniversity Attraction Pole of the Belgian Federal
  Science Policy P5/36. We would like to thank our colleagues Prof.\
  P.\ Dierckx and Prof.\ R.\ Piessens from the Department of 
  Computer Science (KULeuven) for their useful advice. 
\end{acknowledgements} 

%\bibliographystyle{aa}
%\bibliography{5230bib1,5230bib2}

\end{document}